\documentclass[a4paper,10pt,fleqn]{article}
\pdfoutput=1
\usepackage{jheppub}
\usepackage[T1]{fontenc}
\usepackage{multirow, multicol}
\usepackage{braket}
\usepackage{setspace}
\usepackage{enumitem}
\usepackage{amsmath}
\usepackage{amsthm}
\usepackage{bm}
\usepackage{mathtools}
\usepackage{caption}
\usepackage{slashed}
\usepackage{subcaption}
\usepackage{tikz}
\usepackage{lscape}
\usepackage{sidecap}
\usepackage{booktabs}

\definecolor{jlab_red}{RGB}{192,39,45}
\definecolor{jlab_orange}{RGB}{249,102,0}
\definecolor{jlab_blue}{RGB}{47,122,121}
\definecolor{jlab_green}{RGB}{65,125,10}

\hypersetup{%
pdftitle = {ThreePionsIsoImp},
pdfsubject = {QCD},
pdfauthor = {Hansen et al.},
colorlinks = {true},
filecolor = {black},
linkcolor = {jlab_blue},
menucolor = {black},
citecolor = {jlab_green},
urlcolor = {jlab_green},
}{}

\newcommand{\Sec}[0]{section}

\newcommand{\Eq}[0]{eq.}
\newcommand{\Eqs}[0]{eqs.}
\newcommand{\Fig}[0]{figure}
\newcommand{\Reference}[0]{ref.}

\newcommand{\ampyl}[0]{{\sf ampyL}}

\newcommand{\AllTwoBody}[0]{Rummukainen:1995vs,He:2005ey,Christ:2005gi,Kim:2005gf,Lage:2009zv,Bernard:2010fp,Fu:2011xz,Briceno:2012yi,Hansen:2012tf,Guo:2012hv,Briceno:2014oea}

\newcommand{\AllThreeBodyIdenticalSpinZero}[0]{Polejaeva:2012ut,Hansen:2014eka,Hansen:2015zga,Briceno:2017tce,Guo:2017ism,Hammer:2017uqm,Hammer:2017kms,Mai:2017bge,Doring:2018xxx,Briceno:2018mlh,Klos:2018sen,Briceno:2018aml,Guo:2018ibd,Blanton:2019igq,Pang:2019dfe,Romero-Lopez:2019qrt,Blanton:2020gha,Blanton:2020jnm,Guo:2020spn,Romero-Lopez:2020rdq,Muller:2020vtt,Muller:2020wjo,Muller:2021uur,Muller:2022oyw,Jackura:2022xml,Baeza-Ballesteros:2023ljl}

\newcommand{\AllThreeBodyNotISZ}[0]{Hansen:2020zhy,Pang:2020pkl,Blanton:2020gmf,Hansen:2021ofl,Blanton:2021mih,Blanton:2021eyf,Severt:2022jtg,Draper:2023xvu,Bubna:2023oxo}

\newcommand{\GmTwo}[0]{Hoferichter:2019mqg}
\newcommand{\BKKpi}[0]{LHCb:2013lcl,LHCb:2014mir}

\newcommand{\AllLeftHandCut}[0]{Sato:2007ms,Green:2021qol,Padmanath:2022cvl,Meng:2021uhz,Du:2023hlu,Dawid:2023jrj,Raposo:2023oru, Raposo:2025dkb}

\title{Implementing the finite-volume three-pion scattering formalism across all non-maximal isospins}
\author[a,b]{A.~Alotaibi,}
\author[a]{M.~T.~Hansen,}
\author[c,d]{and R.~A.~Brice\~no}
\affiliation[a]{Higgs Centre for Theoretical Physics, School of Physics and Astronomy, The University of Edinburgh, Edinburgh EH9 3FD, UK}
\affiliation[b]{Department of Physics and Astronomy, King Saud University, P.O. Box 2455, Riyadh 11451, Saudi Arabia}
\affiliation[c]{Department of Physics, University of California, Berkeley, CA 94720, USA}
\affiliation[d]{Nuclear Science Division, Lawrence Berkeley National Laboratory, Berkeley, CA 94720, USA}

\emailAdd{a.alotaibi@sms.ed.ac.uk}
\emailAdd{maxwell.hansen@ed.ac.uk}
\emailAdd{rbriceno@berkeley.edu}

\abstract{We present a numerical exploration of the relativistic-field-theory (RFT) formalism for three pions with all possible values of non-maximal isospin, $I_{\pi\pi\pi} = 2$, $1$ and $0$. Using the generic-isospin extension of the RFT formalism \cite{Hansen:2020zhy} and applying our open-source Python library to implement the framework, we predict a range of three-pion energies for illustrative values of the two-to-two scattering amplitudes for various finite-volume irreps also with non-zero total momentum $\boldsymbol P$ in the finite-volume frame. The results restrict attention to the case of a vanishing intrinsic three-body interaction so that the spectra can be understood as a baseline. In future lattice QCD calculations, deviations from these values will be translated into evidence for intrinsic three-body effects in the various scattering channels.}

\begin{document}

\maketitle
\flushbottom
\abovedisplayskip 11pt
\belowdisplayskip 11pt
\clearpage

\section{Introduction}

Within the overwhelmingly successful modern description of fundamental particles known as the Standard Model, the sector that presents the biggest practical challenges for theoretically predicting experimental observables is quantum chromodynamics (QCD), the theory of the strong force. This is because, although the theory is mostly simply expressed in terms of quarks and gluons, these are not the degrees of freedom relevant for processes involving energy or momentum scales at or below $\Lambda_{\sf QCD} \approx 200 \, \text{MeV}$. At such energies, only bound states of quarks and gluons, called hadrons, are observed. Thus, a fundamental challenge is to reliably predict the properties of hadrons from the fundamental theory.

Numerical lattice QCD provides a route to achieve this by numerically evaluating the quantum path integral using Monte Carlo importance sampling. The technique has been established to provide reliable post- and predictions for a range of observables,%
\footnote{See for example the latest Flavour Lattice Averaging Group (FLAG) report~\cite{FlavourLatticeAveragingGroupFLAG:2024oxs}.} provided one can make sense of the modifications to the quantum path integral that are required to make it well-defined and numerically tractable, namely the restriction to a discretized, finite (periodic) spacetime with imaginary (often called Euclidean) time coordinates.

A particularly exciting, cutting-edge application of lattice QCD is in the first-principles determination of multi-hadron scattering amplitudes from finite-volume states. The method requires numerically evaluating various Euclidean two-point correlators with operators drawn from a large set with the same internal quantum numbers. Assembling the two-point functions into a matrix and solving a generalized eigenvalue problem then allows one to numerically estimate the finite-volume energy spectrum of QCD states with periodicity $L$ and quantum numbers dictated by the operators. These energies can then be related to multi-hadron scattering amplitudes through mathematical relations derived from general principles (see refs.~\cite{Briceno:2017max,Hansen:2019nir}, for recent reviews on the topic).

The theoretical formalism to achieve this was first presented in seminal work by L\"uscher \cite{Luscher:1986pf,Luscher:1990ux} for $2\rightarrow 2$ scattering of identical spin-zero particles with zero spatial momentum in the finite-volume frame. This has since been extended by many authors to include non-zero spatial momentum in the finite-volume frame as well as generic two-particle systems with any number of channels, non-identical and non-degenerate particles, and to various possible boundary conditions~\cite{\AllTwoBody}. A limitation of these extensions is that they are valid only in a range of center-of-mass-frame (CMF) energies, up to the lowest lying inelastic (three- or four-particle) threshold.%
\footnote{Interestingly, as has become more relevant over the last few years, the expressions are also only valid above the nearest left-hand cut branchpoint. In the case of $NN \to NN$, for example, this is close to the two-particle threshold, and many calculations see energies on the cut. This issue has recently been addressed in a number of publications~\cite{\AllLeftHandCut}.}

This limited energy reach is one of the key motivations for generalizing such formalisms to include three- (and eventually four-) hadron states. A second key motivation is to rigorously describe resonances that decay into three hadrons, for example, the $\omega$ meson, which is relevant for the anomalous magnetic moment of the muon~\cite{\GmTwo}. A third is to provide a stepping stone to three-particle weak decays, such as $B \to K K \pi$, which are measured by experiments like LHCb to look for new physics via Standard Model precision tests~\cite{\BKKpi}.

By now, significant progress has been made in the finite-volume scattering relations for three-hadron states. While much of this work focuses on the case of three identical spin-zero particles~\cite{\AllThreeBodyIdenticalSpinZero}, recent extensions have also considered different flavor channels, non-degenerate particles, and intrinsic spin~\cite{\AllThreeBodyNotISZ}. In this article, we focus on the generalization of ref.~\cite{Hansen:2020zhy}, which presents the three-particle RFT formalism for three-pion states with all possible isospins. This work aims to provide an implementation of this formalism in a Python library and to explore the phenomenology of the finite-volume energies for all non-maximal isospins of three pions.

To date, the majority of three-hadron lattice QCD calculations have concerned maximal-isospin channels \cite{Mai:2018djl,Horz:2019rrn,Blanton:2019vdk,Culver:2019vvu,Mai:2019fba,Fischer:2020jzp,Hansen:2020otl,Draper:2023boj,Dawid:2025zxc,Dawid:2025doq}.%
\footnote{See, however, refs.~\cite{Mai:2021nul,Sadasivan:2021emk} concerning the $a_1(1260)$ resonance as well as ref.~\cite{Yan:2024gwp} concerning the $\omega(782)$.} Moving away from this restriction is interesting, as all nonmaximal-isospin channels contain two- or three-pion resonances and fully characterizing such resonances is essential to a complete understanding of QCD. These types of calculations are challenging as one has to estimate all possible quark-antiquark pairings (quark contractions) for many operator combinations to reliably extract the energies. Given the scale of the full calculation, it is valuable to break the work down into smaller components.

Here, we focus on the strategy for one such component: relating intermediate-scheme-dependent K-matrices to three-pion finite-volume energies. A second component, which has been the focus of refs.~\cite{Jackura:2020bsk,Dawid:2023jrj, Jackura:2023qtp,Briceno:2024ehy, Jackura:2025wbw}, is to use these K-matrices to extract the physical three-pion scattering amplitudes by solving a set of coupled-integral equations. In particular, refs.~\cite{Jackura:2023qtp,Briceno:2024ehy,Jackura:2025wbw} provide a complete description for studying $3\pi$ scattering amplitudes for all angular momentum and isospin channels.

In this work, we additionally restrict attention to the case of vanishing intrinsic three-body interactions, $\mathcal K_{{\rm df},3} = 0$, where $\mathcal K_{{\rm df},3}$ is the divergence-free K-matrix first introduced in ref.~\cite{Hansen:2014eka}. Thus, this work provides a baseline of predictions for energies in the presence of two-particle interactions. We have taken this restriction for two reasons. First, the finite-volume effects in the presence of a non-zero $\mathcal K_{{\rm df},3}$, are all encoded in the same building blocks as for the $\mathcal K_{{\rm df},3}=0$ case, so that the technical implementation of all $L$-dependent quantities already enters at this stage. Second, there are some subtleties in the implementation of $L$-independent parts of the formalism that arise with non-zero $\mathcal K_{{\rm df},3}$, in particular in the presence of sub-channel resonances. This is discussed in detail in ref.~\cite{Romero-Lopez:2019qrt}.

With this work, we are additionally providing an open-access Python library (called \ampyl) as well as a database for the finite-volume energies corresponding to a range of K-matrix parameters~\cite{ampyL}. We hope this will be useful in benchmarking future calculations and cross-checking and comparing alternative scattering formalisms and codes.

The remainder of this manuscript is structured as follows. In \Sec~\ref{sec:review}, we review the formalism for three-pion scattering in a finite volume. In \Sec~\ref{sec:implementation}, we describe the implementation of the formalism and in \Sec~\ref{sec:results} we present the results of the numerical exploration of the finite-volume energies of three pions for all possible isospins, and in \Sec~\ref{sec:conclusion} we conclude.

\section{RFT formalism with isospin}
\label{sec:review}

Following the original three-particle RFT formalism for identical spin-zero particles~\cite{Hansen:2014eka, Hansen:2015zga} and the extension to include sub-channel resonances~\cite{Briceno:2018aml, Romero-Lopez:2019qrt}, the generalization to all three-pion systems was worked out in ref.~\cite{Hansen:2020zhy}. In this section, we review this formalism in detail before turning to some aspects of its practical implementation.

\subsection{Kinematics}
\label{sec:kinematics}

We restrict attention throughout to three-pion states in a cubic finite volume, with periodicity $L$ in each of the three spatial directions. The system carries a generic total momentum $\boldsymbol{P}$ and total energy $E$ in the finite-volume frame, arranged in the four-vector $P = (E, \boldsymbol P)$. The periodicity of fields implies that all spatial momenta are restricted to integer-vector multiples of $2 \pi/L$. For example
\begin{align}
\boldsymbol{P} = \frac{2\pi}{L}\boldsymbol{d}\,, \qquad {\rm where} \qquad \boldsymbol{d} \in \mathbb{Z}^3 \,,
\end{align}
and we define the CMF energy as $E^\star = \sqrt{E^2 - \boldsymbol{P}^2}$.

Taking two pions within a three-pion-state to carry spatial momentum $\boldsymbol{k}$ and $\boldsymbol{a}$, the corresponding energies are
\begin{align}
\label{eq:omegas}
\omega_k = \sqrt{\boldsymbol{k}^2 + m^2_\pi}\,, \qquad {\rm and } \qquad \omega_a = \sqrt{\boldsymbol{a}^2 + m^2_\pi} \,,
\end{align}
where $m_\pi$ is the physical pion mass, and the on-shell four-momenta are
\begin{align}
k = (\omega_k, \boldsymbol{k})\,, \qquad {\rm and } \qquad a = (\omega_a, \boldsymbol{a}) \,.
\end{align}
We use the mostly-minus metric such that $k^2 = a^2 = m_\pi^2$.
Energy and momentum conservation constrain the third pion to have four-momentum
\begin{equation}
b_{ka} = P-k-a = (E-\omega_k-\omega_a, \boldsymbol{P}-\boldsymbol{k} - \boldsymbol{a}) \,.
\end{equation}
This pion is then only on shell if $\boldsymbol k$ and $\boldsymbol a$ satisfy a constraint such that $b_{ka}^2 = m_\pi^2$. Equivalently, if and only if
\begin{equation}
E - \omega_k - \omega_a = \omega_{ka}\,,
\end{equation}
where
\begin{equation}
\omega_{ka} = \sqrt{(\boldsymbol{P}-\boldsymbol{k} - \boldsymbol{a})^2 + m_\pi^2} \,.
\end{equation}

In the RFT framework, we arrange the three pions as a spectator (with momentum $k$) and a scattering pair or dimer (with momenta $a$ and $b_{ka}$). The dimer can be boosted to its CMF with a boost-velocity $\boldsymbol{\beta_k} = - (\boldsymbol{P}-\boldsymbol{k})/(E-\omega_k)$ and its energy in this frame satisfies
\begin{equation}
E^{\star2}_{2,\boldsymbol{k}} = (E-\omega_k)^2 - (\boldsymbol{P}-\boldsymbol{k})^2 \,,
\end{equation}
where the subscript indicates which momentum is carried by the spectator pion, and the ${}^\star$ generically labels CMF quantities. Applying this boost to $a$ and $b_{ka}$ individually leads to the vectors
\begin{align}
a^\star = (\omega_a^\star, \boldsymbol{a}^\star)\,, \qquad {\rm and } \qquad b_{ka}^\star = (E^\star_{2,\boldsymbol{k}} - \omega_a^\star, -\boldsymbol{a}^\star) \,.
\end{align}
From this expression it is manifest that the $b_{ka}$-pion is on-shell if and only if $\boldsymbol k$ and $\boldsymbol a$ satisfy a constraint such that
\begin{equation}
E_{2,\boldsymbol k}^\star = 2 \omega_a^\star \qquad \Longrightarrow \qquad
q_{2,\boldsymbol{k}}^{\star2} = \boldsymbol a^{\star2} \,,
\end{equation}
where we have defined
\begin{equation}
q_{2,\boldsymbol{k}}^{\star2} = E^{\star 2}_{2,\boldsymbol{k}}/4 - m_\pi^2 \,.
\end{equation}
Therefore, once $P$ and $\boldsymbol{k}$ are fixed, $\hat{\boldsymbol a}^\star$ is the only remaining degree of freedom for the on-shell three-pion state. Such a state can be written as
\begin{equation}
\ket{P,\boldsymbol{k}, \hat{\boldsymbol{a}}^\star} = \sqrt{4\pi} \sum_{\ell m} \ket{P,\boldsymbol{k}, \ell m} Y_{\ell m}(\hat{\boldsymbol{a}}^\star) \,,
\end{equation}
where the spherical harmonics $Y_{\ell m}$ are defined on the unit sphere. In the context of the finite-volume system, $\boldsymbol k$ is also restricted to the finite-volume momenta: $\boldsymbol k = (2 \pi/L) \boldsymbol n$ for $\boldsymbol n \in \mathbb{Z}^3$. Thus, at fixed $P$, one defines a discrete set of allowed states labeled by $\boldsymbol k \ell m$.

This space is discrete but infinite since an infinite tower of angular momenta can contribute. As we discuss more in the following section, in practice this is truncated at some maximum angular momentum $\ell_{\sf max}$. By contrast, the $\boldsymbol k$-space is rendered finite by the kinematics of the system. To see this, note that, for sufficiently large $\boldsymbol{k}^2$, $ E^{\star2}_{2,\boldsymbol{k}}$ is negative and it is not possible for all three pions to go on the mass shell. As originally discussed in ref.~\cite{Luscher:1986pf}, on-shell intermediate states correspond to power-like $L$ dependence and it is only this type of dependence that we aim to control. Thus, we can neglect contributions from sufficiently large $\boldsymbol{k}^2$. This requires introducing a smooth cutoff function, denoted $H(E_{2,\boldsymbol k}^{\star 2})$, to the quantities entering the quantization condition.%
\footnote{We note that the notation here $H(E_{2,\boldsymbol k}^{\star 2})$ differs slightly from ref.~\cite{Hansen:2014eka} where we write $H(\boldsymbol k)$. However, the underlying function is the same as in the earlier publication.} This is defined explicitly in \Sec~\ref{sec:Kmatrix}.

\subsection{Isospin}
\label{sec:isospin}

The next step is to define states with definite isospin. This is performed following the approach detailed in ref.~\cite{Hansen:2020zhy}. First, we recall that the basic decomposition of isospin
\begin{align}
1_\pi \otimes 1_\pi \otimes 1_\pi & = (0_{\pi \pi} \oplus 1_{\pi \pi} \oplus 2_{\pi \pi}) \otimes 1_{\pi} \,,
\\[3pt]
& = (1)_{[I_{\pi \pi} = 0 (\sigma)] } \ \oplus \ (0 \oplus 1 \oplus 2)_{[I_{\pi \pi} = 1 (\rho)]} \ \oplus \ (1 \oplus 2 \oplus 3)_{[I_{\pi \pi} = 2]} \,,
\\[5pt]
\begin{split}
& = \left [ 0_{[I_{\pi \pi} = 0 (\sigma)]} \right ] \ \oplus \ \left [1_{[I_{\pi \pi} = 0 (\sigma)]} \ \oplus \ 1_{[I_{\pi \pi} = 1 (\rho)]} \ \oplus \ 1_{[I_{\pi \pi} = 2]} \right ] \\[2pt]
& \hspace{140pt} \oplus \ [2_{[I_{\pi \pi} = 1 (\rho)]} \ \oplus \ 2_{[I_{\pi \pi} = 2]}] \ \oplus \ [3_{[I_{\pi \pi} = 2]}] \,.
\end{split}
\end{align}
Here, the left-hand side indicates the combination of three pions each with isospin one. The first line indicates an intermediate step in which the first two pions are combined to have definite isospin. In the second line, each value indicates a definite total three-pion isospin, while the subscripts denote the isospin of the two-pion system. This includes the labels in parentheses for the $I_{\pi \pi} = 0$ and $I_{\pi \pi} = 1$ cases, of $(\sigma)$ and $(\rho)$ respectively, to recall the two-pion resonances that are present in these channels. On the final line, we rearrange this information, sorted by the total isospin of the three-pion system.

The same information can be summarized by introducing the state vectors
\begin{align}
\boldsymbol I_3 & = \begin{pmatrix} \ket{(\pi \pi)_2 \pi}_3 \end{pmatrix} \,,
\label{eq:iso3_vector}
\\[5pt]
\boldsymbol I_2 & = \begin{pmatrix} \ket{(\pi \pi)_2 \pi}_2 \\ \ket{\rho \pi}_2 \end{pmatrix} \,,
\\[5pt]
\boldsymbol I_1 & = \begin{pmatrix} \ket{(\pi \pi)_2 \pi}_1 \\ \ket{\rho \pi}_1 \\ \ket{\sigma \pi}_1 \end{pmatrix} \,,
\label{eq:iso1_vector}\\[5pt]
\boldsymbol I_0 & = \begin{pmatrix} \ket{\rho \pi}_0 \end{pmatrix} \,.
\label{eq:iso0_vector}
\end{align}
Each state appearing within the column vectors on the right-hand side here corresponds to a multiplet with $2 I_{\pi \pi \pi}+1$ entries, where $I_{\pi \pi \pi}$ is the total three-pion isospin indicated by the subscript outside the ket. Thus the total number of states is given by $\dim[\boldsymbol I_{3}] = 7$, $\dim[\boldsymbol I_{2}] = 10$, $\dim[\boldsymbol I_{1}] = 9$, $\dim[\boldsymbol I_{0}] = 1$ for a total of $27$ states. This matches the counting in terms of three pions each carrying one of three isospin values (equivalently, electric charges).

In the derivation of ref.~\cite{Hansen:2020zhy}, it was convenient to focus on a specific entry of the multiplet and this is mostly easily achieved by restricting to the charge neutral state. In the basis of individual pions seven combinations are possible, denoted by
\begin{multline}
\boldsymbol V^T = \bigg (
\ket{\pi_+, \pi_-, \pi_0} \,, \ket{\pi_+, \pi_0, \pi_-} \,, \ket{\pi_-, \pi_+, \pi_0} \,, \ket{\pi_-, \pi_0, \pi_+} \,, \\ \ket{\pi_0, \pi_+, \pi_-} \,, \ket{\pi_0, \pi_-, \pi_+} \,, \ket{\pi_0, \pi_0, \pi_0}
\bigg )\,.
\label{eq:V_vector}
\end{multline}

Then it is possible to form linear combinations of the entries of $\boldsymbol V$ to generate the neutral parts of the vectors $\boldsymbol I_3$, $\boldsymbol I_2$, $\boldsymbol I_1$ and $\boldsymbol I_0$. This generically takes the form
\begin{align}
{\boldsymbol I}^{\sf o}_{I_{\pi \pi \pi}} & = C_{I_{\pi \pi \pi}} \cdot \boldsymbol V \,,
\label{eq:transform}
\end{align}
where the ${}^{\sf o}$ superscript indicates restriction to the neutral states and $C_{I_{\pi \pi \pi}}$ is a $\text{dim}[\boldsymbol I^{\sf o}_{I_{\pi \pi \pi}}] \times 7$ matrix. The explicit form of $C_{I_{\pi \pi \pi}}$ is given in eqs.~(2.56) and (2.57) of \Reference~\cite{Hansen:2020zhy}.

The result of this construction is that four quantization conditions are derived, one each for the four possible values of $I_{\pi \pi \pi}$. Each of these is defined via matrices with indices both in the $\boldsymbol k \ell m$ space and in a flavor space $f$ resulting from the two-pion isospin values. The dimensionality of each flavor space can be read off of the $\boldsymbol I_{I_{\pi \pi \pi}}$ vectors in \Eqs~\eqref{eq:iso3_vector}-\eqref{eq:iso0_vector}. Since only the neutral states are considered, each entry within the column vectors is counted once, leading to $1,2,3,1$ for $I_{\pi \pi \pi} = 3,2,1,0$ respectively. We denote the full index space via $f \boldsymbol k \ell m$.

We now turn to the three relevant matrices required to specify the $\mathcal K_{{\rm df},3} = 0$ quantization conditions. These are each defined on the $f \boldsymbol k \ell m$ space and are denoted by $\textbf K^{[I_{\pi \pi \pi}]}$ (encoding subprocess two-to-two interactions), $\textbf F^{[I_{\pi \pi \pi}]}$ (encoding finite-volume cuts with the same spectator on either side of the cut), and $\textbf G^{[I_{\pi \pi \pi}]}$ (encoding cuts with different spectators on either side).

\subsection{K-matrix}
\label{sec:Kmatrix}

First, making the flavor space explicit, we write
\begin{align}
\textbf K^{[3]} & =
\begin{pmatrix}
P^{\{2\} \dagger} \mathcal K^{\{2\}} P^{\{2\}}
\end{pmatrix} \,,
\\
\textbf K^{[2]} & =
\begin{pmatrix}
P^{\{2\} \dagger} \mathcal K^{\{2\}} P^{\{2\}} & 0 \\
0 & P^{\{1\} \dagger} \mathcal K^{\{1\}} P^{\{1\}}
\end{pmatrix} \,,
\\
\textbf K^{[1]} & =
\begin{pmatrix}
P^{\{2\} \dagger} \mathcal K^{\{2\}} P^{\{2\}} & 0 & 0 \\
0 & P^{\{1\} \dagger} \mathcal K^{\{1\}} P^{\{1\}} & 0 \\
0 & 0 & P^{\{0\} \dagger} \mathcal K^{\{0\}} P^{\{0\}}
\end{pmatrix} \,,
\\
\textbf K^{[0]} & =
\begin{pmatrix}
P^{\{1\} \dagger} \mathcal K^{\{1\}} P^{\{1\}}
\end{pmatrix} \,,
\end{align}
where we have adopted the notation that $\mathcal K^{\{I_{\pi \pi}\}}$ is a K-matrix block with definite two-pion isospin $I_{\pi \pi} = 2,1,0$. Note that for each value of $I_{\pi \pi \pi}$, the K-matrices are populated with all values of $I_{\pi \pi}$ that satisfy the isospin decomposition given in \Eqs~\eqref{eq:iso3_vector}-\eqref{eq:iso0_vector}.

We have also introduced $P^{\{I_{\pi \pi}\}}$, which are matrices on the $\boldsymbol k \ell m$ space that restrict the K-matrices according to the fact that odd (even) two-pion isospins can only have odd (even) angular momenta, while also enforcing an angular momentum truncation described in more detail below. To be concrete, we take $\mathcal K^{\{I_{\pi \pi}\}}$ to formally be populated by all angular momenta (with zeroes for the even or odd entries, depending on $I_{\pi \pi}$). $P^{\{I_{\pi \pi}\}}$ then discards the trivial even set. As a particular example, for $I_{\pi \pi}=1$ a natural truncation is to choose $\ell_{\sf max} = 1$, leading to
\begin{equation}
P_{\boldsymbol k' \ell' m', \boldsymbol k \ell m}^{\{1\}} = \delta_{\boldsymbol k' \boldsymbol k} \begin{pmatrix}
0 & 0 & 0 \\ 1 & 0 & 0 \\ 0 & 1 & 0 \\ 0 & 0 & 1 \\ 0 & 0 & 0 \\ \vdots & \vdots & \vdots \end{pmatrix}_{\ell'm',\ell m} \,. \label{eq:P1}
\end{equation}
This has the effect of discarding everything but the $\ell = 1$ entries in the isospin-one block of the K-matrix.

The remaining undefined quantity, $\mathcal K^{\{I_{\pi \pi}\}}$, is a matrix on the $\boldsymbol k \ell m$ space with matrix elements
\begin{equation}
\mathcal K^{\{I_{\pi \pi}\}}_{\boldsymbol k' \ell' m', \boldsymbol k \ell m} = \frac{1}{2 \omega_k} \delta_{\boldsymbol k' \boldsymbol k} \delta_{\ell' \ell} \delta_{m' m} \mathcal K^{\{I_{\pi \pi}\}}_{\ell}(E_{2, \boldsymbol k}^{\star 2}) \,,
\end{equation}
where $\mathcal K^{\{I_{\pi \pi}\}}_{\ell}(E_{2, \boldsymbol k}^{\star 2})$ is a simple real function for each $I_{\pi \pi}$ and $\ell$, with no implicit indices. It is related to scattering phase shift by
\begin{equation}
\mathcal K^{\{I_{\pi \pi}\}}_{\ell}(E_{2,\boldsymbol k}^{\star 2})^{-1} = p^{2 \ell} \frac{p \cot \delta^{\{I_{\pi \pi}\}}_{\ell}(p) + \vert p \vert [1 - H(E_{2,\boldsymbol k}^{\star 2})]}{16 \pi \sqrt{E_{2,\boldsymbol k}^{\star 2}}} \bigg \vert_{p = \sqrt{E_{2,\boldsymbol k}^{\star 2}/4 - m_\pi^2}} \,.
\end{equation}
This K-matrix is unconventional in two ways. First, we have explicitly canceled the $p^{2\ell}$ barrier factors. Second, this relation depends on the cutoff function $H(E_{2,\boldsymbol k}^{\star 2})$ referenced in the previous section. Various choices are possible, but for the majority of this work we use the choice first described in ref.~\cite{Briceno:2017tce}, which is a variant on the original choice defined in ref.~\cite{Hansen:2014eka}. We write
\begin{align}
H(E_{2,\boldsymbol k}^{\star 2}) & \equiv J(z) \bigg \vert_{z = \mathcal B_{\alpha}(E_{2,\boldsymbol k}^{\star 2}/(4 m_\pi^2))} \,,
\label{eq:Hdef}
\end{align}
where
\begin{align}
& J(z) \equiv
\begin{dcases}
0, & 0 \geq z\,;\\
\exp\bigg(-\frac{1}{z}\exp\bigg[-\frac{1}{1-z}\bigg]\bigg), & 0 < z < 1 \,; \\
1, & 1 \leq z\,,
\end{dcases}
\label{eq:smoothcutoff}
\\[10pt]
& \mathcal B_{\alpha}(E_{2,\boldsymbol k}^{\star 2}/(4 m_\pi^2)) \equiv \frac{E_{2,\boldsymbol k}^{\star 2}/(4 m_\pi^2) - (1+\alpha)/4 }{ (3-\alpha)/4 }.
\label{eq:z_alpha}
\end{align}
One recovers the choice of ref.~\cite{Hansen:2014eka} by setting $\alpha = -1$:
\begin{equation}
\mathcal B_{-1}(E_{2,\boldsymbol k}^{\star 2}/(4 m_\pi^2)) \equiv E_{2,\boldsymbol k}^{\star 2}/(4 m_\pi^2) \,.
\end{equation}
The advantage of having a more flexible definition is that it provides a simple dial as to where to enforce the cutoff function to vanish exactly. In particular, we see that for any $\alpha$, $H(E_{2,\boldsymbol k}^{\star 2})$ has zero support for $E_{2,\boldsymbol k}^{\star 2} < (1+\alpha) \, m_\pi^2 $.

$\alpha$ must be chosen to cut off the contribution of subthreshold singularities in the two-body K matrices that are either unphysical but present in the parametrization, or else physical but absent in the parameterization used. In this work, we consider examples of the former when using a Breit--Wigner description. This leads to deep unphysical poles, and if $\alpha$ is chosen such that the unphysical poles contribute, this contributes unphysical behavior in the predicted finite-volume energy spectrum.

At this stage, all that is needed to evaluate $\textbf K^{[I_{\pi \pi \pi}]}$ numerically is a set of truncations $\ell_{\sf max}^{\{I_{\pi \pi}\}}$ for each isospin channel together with a set of phaseshift parametrizations. We represent the latter by
\begin{equation}
\delta^{{\sf par} ,\{I_{\pi \pi}\}}_{\ell}(p \, \vert \, \boldsymbol \eta^{par,\{I_{\pi \pi}\}}_{\ell})\,,
\end{equation}
where ${\sf par}$ stands for a particular phaseshift parametrization and $\boldsymbol \eta^{{\sf par},\{I_{\pi \pi}\}}_{\ell}$ is a vector of the parameters. We give details of the particular choices used in this work in \Sec~\ref{sec:Kmatrix_param}.

We now turn to the finite-volume cuts entering the quantization condition, beginning with $\textbf F^{[I_{\pi \pi \pi}]}$.

\subsection{F-matrix}

In direct imitation to the steps above, we begin by making the flavor space explicit by writing
\begin{align}
\textbf F^{[3]} & =
\begin{pmatrix}
P^{\{2\} \dagger} F P^{\{2\}}
\end{pmatrix} \,,
\\
\textbf F^{[2]} & =
\begin{pmatrix}
P^{\{2\} \dagger} F P^{\{2\}} & 0 \\
0 & P^{\{1\} \dagger} F P^{\{1\}}
\end{pmatrix} \,,
\\
\textbf F^{[1]} & =
\begin{pmatrix}
P^{\{2\} \dagger} F P^{\{2\}} & 0 & 0 \\
0 & P^{\{1\} \dagger} F P^{\{1\}} & 0 \\
0 & 0 & P^{\{0\} \dagger} F P^{\{0\}}
\end{pmatrix} \,,
\\
\textbf F^{[0]} & =
\begin{pmatrix}
P^{\{1\} \dagger} F P^{\{1\}}
\end{pmatrix} \,.
\end{align}
Here, we are making use of the fact that the $P^{\{I_{\pi \pi}\}}$ are defined for all $\ell m$ and can thus be applied to the same fundamental quantity.

The remaining matrix $F$ is defined on the $\boldsymbol k \ell m$ space with matrix elements
\begin{align}
\label{eq:F_kellm_space}
F_{\boldsymbol k'\ell'm',\boldsymbol k\ell m} & \equiv \delta_{\boldsymbol k' \boldsymbol k} F_{\ell' m'\ell m}(\boldsymbol{k})\,, \\
F_{\ell'm',\ell m}(\boldsymbol{k} ) & \equiv \frac{H(E_{2,\boldsymbol{k}}^\star)}{4 \omega_k} \bigg[\frac{1}{L^3} \sum_{\boldsymbol{a}} - {\rm pv}\int\frac{d^3 \boldsymbol a}{(2\pi)^3} \bigg] \frac{4\pi Y_{\ell'm'}(\hat{\boldsymbol{a}}^\star)Y^*_{\ell m}(\hat{\boldsymbol{a}}^\star) \vert \boldsymbol{a}^\star \vert^{\ell'+\ell} e^{- \alpha(\boldsymbol a^{\star 2}-q_{2,k}^{\star2})}}{2\omega_a 2\omega_{ka}(E-\omega_k-\omega_a-\omega_{ka})} \,,
\label{eq:F_pv}
\end{align}
where the principal value integral is defined as the real part of the $i \epsilon$ pole prescription. Here we have included the UV regulator $e^{- \alpha(\boldsymbol a^{\star 2}-q_{2,k}^{\star2})}$. Following ref.~\cite{Kim:2005gf}, it is understood that $\alpha \to 0^+$ should be estimated to eliminate any enhanced exponentially suppressed volume effects arising from the regulator.

To implement this in practice, we use the form
\begin{align}
F_{\ell'm',\ell m}(\boldsymbol{k} ) & = \frac{H(E_{2,\boldsymbol{k}}^\star)}{8 \omega_k} \left [ S_{\ell'm',\ell m}(\boldsymbol{k} ) - I_{\ell'm',\ell m}(\boldsymbol{k} )\right ] \,,
\\
S_{\ell'm',\ell m}(\boldsymbol{k} ) & \equiv \frac{1}{L^3} \sum_{\boldsymbol{a}} \frac{4\pi Y_{\ell'm'}(\hat{\boldsymbol{a}}^\star)Y^*_{\ell m}(\hat{\boldsymbol{a}}^\star) \vert \boldsymbol{a}^\star \vert^{\ell'+\ell} e^{- \alpha(\boldsymbol a^{\star 2}-q_{2,k}^{\star2})}}{2\omega_a (q_{2, \boldsymbol k}^{\star 2} - a^{\star2})} \,, \\
I_{\ell'm',\ell m}(\boldsymbol{k} ) & \equiv \delta_{\ell \ell'} \delta_{mm'} {\rm pv}\int\frac{d^3 \boldsymbol a^\star}{(2\pi)^3} \frac{ \vert \boldsymbol{a}^\star \vert^{\ell'+\ell} e^{- \alpha(\boldsymbol a^{\star 2}-q_{2,k}^{\star2})}}{2\omega_a^\star (q_{2, \boldsymbol k}^{\star 2} - a^{\star2})} \,,
\end{align}
which is equivalent (up to exponentially suppressed volume effects) as was shown in refs.~\cite{Kim:2005gf,Briceno:2018mlh}.

\subsection{G-matrix}

Finally, we reach $\textbf G^{[I_{\pi \pi \pi}]}$, whose flavor-space definition is the most complicated of the three quantities. The required expression can be given compactly as
\begin{align}
\textbf G^{[3]} & = \begin{pmatrix} 1 \end{pmatrix}
\circ P^{\{2\} \dagger} GP^{\{2\}}\,,
\\
\textbf G^{[2]} & = \begin{pmatrix}
-\frac{1}{2} & -\frac{\sqrt{3}}{2} \\
\frac{\sqrt{3}}{2} & -\frac{1}{2}
\end{pmatrix}
\circ
\begin{pmatrix}
P^{\{2\} \dagger} \\ P^{\{1\} \dagger}
\end{pmatrix}
G
\begin{pmatrix}
P^{\{2\}} & P^{\{1\}}
\end{pmatrix}\,,
\\
\textbf G^{[1]} & = \begin{pmatrix} \frac{1}{6} & \frac{\sqrt{15}}{6} & \frac{\sqrt{5}}{3} \\
\frac{\sqrt{15}}{6} & \frac{1}{2} & -\frac{1}{\sqrt{3}} \\
\frac{\sqrt{5}}{3} & -\frac{1}{\sqrt{3}} & \frac{1}{3} \end{pmatrix}
\circ
\begin{pmatrix}
P^{\{2\} \dagger} \\ P^{\{1\} \dagger} \\ P^{\{0\} \dagger}
\end{pmatrix}
G
\begin{pmatrix}
P^{\{2\}} & P^{\{1\}} & P^{\{0\}}
\end{pmatrix}\,,
\\
\textbf G^{[0]} & = \begin{pmatrix} -1 \end{pmatrix}
\circ P^{\{1\} \dagger} GP^{\{1\}}\,.
\end{align}
Here, $\circ$ represents the Hadamard product (elementwise matrix multiplication).
The $P^{\{I_{\pi \pi}\}}$ project the matrix $G$ onto the same $\ell m$ space as $\textbf K^{[I_{\pi \pi \pi}]}$ and $\textbf F^{[I_{\pi \pi \pi}]}$, but now the column and row vectors are required to represent the various projections arising in the off-diagonal elements. Each entry of the projected matrix is then multiplied by the appropriate factor from the Hadamard product to reflect the isospin projection as detailed in ref.~\cite{Hansen:2020zhy}.

The underlying $G$ matrix is defined on the $\boldsymbol k \ell m$ space with matrix elements
\begin{equation}
\label{G_function}
G_{\boldsymbol p \ell' m' , \boldsymbol k \ell m} \equiv \frac{1}{2\omega_k 2 \omega_p L^3} \frac{4\pi Y_{\ell'm'}(\hat{\boldsymbol{k}}^\star)\vert \boldsymbol{k}^\star\vert^{\ell'} Y^*_{\ell m}(\hat{\boldsymbol{p}}^\star) \vert \boldsymbol{p}^\star\vert^\ell}{ b_{kp}^2 - m_\pi^2 } H(E_{2,\boldsymbol{k}}^{\star 2}) H(E_{2,\boldsymbol{p}}^{\star 2}) \,.
\end{equation}
Here, $\boldsymbol{k}^\star$ is the spatial component of $k^{\star} = (\omega_k^\star, \boldsymbol{k}^\star)$, defined by boosting $k=(\omega_k, \boldsymbol{k})$ with boost velocity $\boldsymbol{\beta_p} = - (\boldsymbol{P} - \boldsymbol{p})/(E - \omega_p)$. The same is true for $\boldsymbol{p}^\star$ with $k \leftrightarrow p$.

\subsection{Quantization condition}
\label{sec:QC}

Having constructed the three matrices $\textbf K^{[I_{\pi \pi \pi]}}, \textbf F^{[I_{\pi \pi \pi]}},$ and $\textbf G^{[I_{\pi \pi \pi}]}$, it is now straightforward to write down the $\mathcal K_{{\rm df},3}=0$ quantization condition for the three-pion system~\cite{Hansen:2020zhy}. We first define
\begin{equation}
\label{eq:QC_3}
\mathcal Q^{[I_{\pi \pi \pi}]}(E, \boldsymbol P, L \, \vert \, \boldsymbol \eta) = {\rm det}_{f \boldsymbol k \ell m} \left [ 1+ (\textbf F^{[I_{\pi \pi \pi}]} + \textbf G^{[I_{\pi \pi \pi}]}) \textbf K^{[I_{\pi \pi \pi}]} \right ] \,.
\end{equation}
As emphasized by the functional dependence on the left-hand side, this determinant is a function of the energy $E$, the total momentum $\boldsymbol P$, the volume $L$, and the parameters $\boldsymbol \eta$ that define the K-matrix.

The quantization condition is then the statement that, provided $E^\star = \sqrt{E^2 - \boldsymbol P^2} < 5 m_\pi$ (and up to neglected exponentially suppressed $L$-dependence), this determinant vanishes at the finite-volume energies $E_n^{[I_{\pi \pi \pi}]}(L, \boldsymbol P, \boldsymbol \eta)$:%
\footnote{In the presence of even-odd coupling, the range of limited validity would be set by a four-particle rather than a five-particle state.}
\begin{equation}
\mathcal Q^{[I_{\pi \pi \pi}]}(E_n^{[I_{\pi \pi \pi}]}(L, \boldsymbol P, \boldsymbol \eta), \boldsymbol P, L \, \vert \, \boldsymbol \eta) = 0 \,.
\end{equation}
This is the central result of the formalism when the three-body K-matrix is fixed to zero.

\section{Implementation}
\label{sec:implementation}

We turn now to the implementation of the quantization condition to extract the finite-volume energies of three-pion states with non-maximal isospin: $I_{\pi\pi\pi} = 2,1$, and $0$.

\subsection{Finite-volume group theory}
\label{sec:fv_group}

In lattice QCD calculations, it is advantageous (and standard) to extract finite-volume energies from correlation functions with definite values for all quantum numbers that are well-defined in the finite-volume system. This partially mitigates the challenge of reliably determining the energies from fits to Euclidean correlation functions.

In the context of the present work, this means that finite-volume spectra are labelled with their value of three-pion isospin $I_{\pi \pi \pi}$ and with the total momentum $\boldsymbol P$ in the finite-volume frame. The value of the latter, together with the cubic, periodic volume, defines a finite symmetry group and the irreducible representations (irreps) of the latter provide an additional label. This work restricts attention to $\boldsymbol P = [000],[001],[011]$. The names and basic properties of the corresponding symmetry groups are summarized in table~\ref{tab:irrep_review}.

\begingroup
\setlength{\tabcolsep}{8pt}
\renewcommand{\arraystretch}{1.2}
\begin{table}[t]
\centering
\begin{tabular}{c|c|c}
\hline\hline
$\boldsymbol{P}$ & $G\ (\vert G \vert)$ & irrep (dimension) \\
\hline\hline
\multirow{5}{*}{$[000]$} & \multirow{5}{*}{$O_h\ (48)$} & $A^{\pm}_1\ (1)$ \\
& & $A_2^{\pm}\ (1)$ \\
& & $E^{\pm}\ (2)$\\
& & $T_1^{\pm}\ (3)$ \\
& & $T_2^{\pm}\ (3)$\\
\hline
\multirow{5}{*}{$[001]$} & \multirow{5}{*}{${\rm Dic}_4\ (8)$} & $A_1\ (1)$\\
& & $A_2\ (1)$\\
& & $B_1\ (1)$\\
& & $B_2\ (1)$ \\
& & $E_2\ (2)$\\
\hline
\multirow{4}{*}{$[011]$} & \multirow{4}{*}{${\rm Dic}_2\ (4)$} & $A_1\ (1)$\\
& & $A_2\ (1)$\\
& & $B_1\ (1)$\\
& & $B_2\ (1)$ \\
\hline\hline
\end{tabular}
\caption{Summary of the irreducible representations (irreps) of the finite-volume symmetry group for a given total momentum $\boldsymbol{P}$. The value of $\boldsymbol P$ is given using the shorthand $\boldsymbol P = [d_xd_yd_z]$ for $\boldsymbol P = (2 \pi/L)(d_x, d_y, d_z)$. The second column gives the name of the group and the number of elements, while the third gives the name of the irrep and its dimension. Note for $\boldsymbol P = [000]$ there are two copies of each irrep, one for each parity as indicated by the superscript $\pm$, leading to 10 irreps in total.}
\label{tab:irrep_review}
\end{table}
\renewcommand{\arraystretch}{1}
\endgroup

To predict finite-volume energies in definite irreps, it is necessary to project the matrices in \Eq~\eqref{eq:QC_3} before evaluating the determinant. As a warm-up to this, and also as it is useful in its own right, we begin by considering how the non-interacting three-pion states split across the irreps. First, note that a generic, non-interacting three-pion energy can be written as
\begin{align}
E^{\{\boldsymbol n\}}(L) = \sum_{i=1}^{3} \sqrt{m_\pi^2 + (2\pi/L)^2 \boldsymbol n^2_i} \,,
\label{eq:non_interacting_energy}
\end{align}
where $\{\boldsymbol n\} = \{\boldsymbol n_1\boldsymbol n_2\boldsymbol n_3\}$ is an array populated with dimensionless momenta of the three individual pions. Because we will always consider states with definite total momentum, $\{\boldsymbol n\}$ will be constrained to satisfy
\begin{equation}
\sum_{i=1}^{3} \boldsymbol n_i = L \boldsymbol{P} / (2\pi) \,.
\end{equation}
We will also find it useful to define the notation $S(\boldsymbol n) = \{\boldsymbol n^2_1 \boldsymbol n^2_2 \boldsymbol n^2_3\}$, where $\boldsymbol n^2_1$, $ \boldsymbol n^2_2$, $ \boldsymbol n^2_3$ are chosen such that the integers populating $S(\boldsymbol n)$ are ascending. So, for example, in the case of $\boldsymbol P = \boldsymbol 0$, the $S(\boldsymbol n)$ values for the lowest four energy multiplets are $\{000\}$, $\{011\}$, $\{022\}$ and $\{112\}$. Each of these corresponds to a number of distinct pion states with definite individual pion flavors. Focusing on the neutral sector, both three identical neutral pions ($\pi_0, \pi_0, \pi_0$) and three distinct pions ($\pi_+, \pi_-,\pi_0$) are relevant. In the second and third columns of tables~\ref{tab:counting_ni_pipipi_P000},~\ref{tab:counting_ni_pipipi_P001}, and~\ref{tab:counting_ni_pipipi_P011} below, we summarise the number of non-interacting states with either three identical or three non-identical pions (for each $S(\boldsymbol n)$ value) for $\boldsymbol P = [000], [001],[011]$, respectively.%
\footnote{All tables related to the non-interacting states and their decomposition into definite isospin and finite-volume irreducible representations were generated using the \ampyl\ library. We gratefully acknowledge Christopher Thomas for assistance with various cross-checks of this code.}

Note that the useful definition for a set of states is in fact all those that are related by swaps of the individual momenta together with rotations taken from the little group for a given $\boldsymbol P$. While the $S(\boldsymbol n)$ label is often sufficient to identify such a set, this is not always the case. In such cases, for example for $\boldsymbol P = [001]$ and $\{111\}$, one needs to specify the individual momenta as we have done in the caption of table~\ref{tab:counting_ni_pipipi_P001}.

\begingroup
\setlength{\tabcolsep}{8pt}
\renewcommand{\arraystretch}{1.2}
\begin{table}[t]
\centering
$\boldsymbol P = [000]$
\vspace{10pt}

\begin{tabular}{c|c|c || c | c | c | c}
\hline\hline
$S(\boldsymbol n)$ & $N_{\pi_0 \pi_0 \pi_0}$ & $N_{\pi_+ \pi_- \pi_0}$ & $I_{\pi \pi \pi} = 3$& $I_{\pi \pi \pi} = 2$& $I_{\pi \pi \pi} = 1$& $I_{\pi \pi \pi} = 0$ \\
\hline\hline
\{000\} & $1$ & $1$ & $1$ & $0$ & $1$ & $0$\\
\{011\} & $3$ & $18$ & $3$ & $6$ & $9$ & $3$\\
\{022\} & $6$ & $36$ & $6$ & $12$ & $18$ & $6$\\
\{112\} & $12$ & $72$ & $12$ & $24$ & $36$ & $12$\\
\{033\} & $4$ & $24$ & $4$ & $8$ & $12$ & $4$\\
\hline\hline
\end{tabular}
\caption{Summary of the number of possible $\boldsymbol{P} = [000]$ non-interacting states within a given set $S(\boldsymbol{n})$, for systems of either three identical pions, three non-identical pions, or with a definite three-pion isospin. The first column lists the momentum type, using the notation $S(\boldsymbol{n}) = \{\boldsymbol{n}_1^2\, \boldsymbol{n}_2^2\, \boldsymbol{n}_3^2\}$ for a state whose energy is given by \Eq~\eqref{eq:non_interacting_energy}. For each row, the sum of the entries in the second and third columns equals the sum of the entries in the fourth through seventh columns.}
\label{tab:counting_ni_pipipi_P000}
\end{table}
\renewcommand{\arraystretch}{1}
\endgroup

\begingroup
\setlength{\tabcolsep}{8pt}
\renewcommand{\arraystretch}{1.2}
\begin{table}[t]
\centering
$\boldsymbol P = [001]$
\vspace{10pt}

\begin{tabular}{c|c|c || c | c | c | c}
\hline\hline
$S(\boldsymbol n)$ & $N_{\pi_0 \pi_0 \pi_0}$ & $N_{\pi_+ \pi_- \pi_0}$ & $I_{\pi \pi \pi} = 3$& $I_{\pi \pi \pi} = 2$& $I_{\pi \pi \pi} = 1$& $I_{\pi \pi \pi} = 0$ \\
\hline\hline
\{001\} & $1$ & $3$ & $1$ & $1$ & $2$ & ---\\
\{012\} & $4$ & $24$ & $4$ & $8$ & $12$ & $4$\\
\{111\} & $2$ & $12$ & $2$ & $4$ & $6$ & $2$\\
$\{111\}^*$ & $1$ & $3$ & $1$ & $1$ & $2$ & ---\\
\{014\} & $1$ & $6$ & $1$ & $2$ & $3$ & $1$\\
\{023\} & $4$ & $24$ & $4$ & $8$ & $12$ & $4$\\
\{113\} & $4$ & $24$ & $4$ & $8$ & $12$ & $4$\\
\hline\hline
\end{tabular}
\caption{Summary of the number of possible $\boldsymbol{P} = [001]$ non-interacting states, with other details as in table~\ref{tab:counting_ni_pipipi_P000}. In this case, the label $S(\boldsymbol{n}) = \{111\}$ is insufficient, as two distinct momentum sets exist that are not related by rotations leaving $\boldsymbol{P}$ invariant. We define $\{111\}$ to denote rotations and permutations of $\boldsymbol{n}_1 = (0,1,0)$, $\boldsymbol{n}_2 = (0,-1,0)$, $\boldsymbol{n}_3 = (0,0,1)$, while $\{111\}^*$ denotes rotations and permutations of $\boldsymbol{n}_1 = (0,0,1)$, $\boldsymbol{n}_2 = (0,0,1)$, $\boldsymbol{n}_3 = (0,0,-1)$.}
\label{tab:counting_ni_pipipi_P001}
\end{table}
\renewcommand{\arraystretch}{1}
\endgroup

\begingroup
\setlength{\tabcolsep}{8pt}
\renewcommand{\arraystretch}{1.2}
\begin{table}[t]
\centering
$\boldsymbol P = [011]$
\vspace{10pt}

\begin{tabular}{c|c|c || c | c | c | c}
\hline\hline
$S(\boldsymbol n)$ & $N_{\pi_0 \pi_0 \pi_0}$ & $N_{\pi_+ \pi_- \pi_0}$ & $I_{\pi \pi \pi} = 3$& $I_{\pi \pi \pi} = 2$& $I_{\pi \pi \pi} = 1$& $I_{\pi \pi \pi} = 0$ \\
\hline\hline
\{002\} & $1$ & $3$ & $1$ & $1$ & $2$ & ---\\
\{011\} & $1$ & $6$ & $1$ & $2$ & $3$ & $1$\\
\{012\} & $2$ & $12$ & $2$ & $4$ & $6$ & $2$\\
\{022\} & $2$ & $12$ & $2$ & $4$ & $6$ & $2$\\
\{112\} & $4$ & $24$ & $4$ & $8$ & $12$ & $4$\\
\hline\hline
\end{tabular}
\caption{Summary of the number of possible $\boldsymbol P=[011]$ non-interacting states, with other details as in table~\ref{tab:counting_ni_pipipi_P000}.}
\label{tab:counting_ni_pipipi_P011}
\end{table}
\renewcommand{\arraystretch}{1}
\endgroup

The next step is to determine how these multiplicities divide across the four isospin sectors. This is achieved by taking the set of definite isospin states, given by \Eqs~\eqref{eq:iso3_vector} to \eqref{eq:transform}, assigning all permutations and rotations of $\boldsymbol n_1, \boldsymbol n_2,$ and $\boldsymbol n_3$ from a given set, and then counting the number distinct, non-zero states that appear. For example, in the case of $\boldsymbol P=[000]$ and $S(\boldsymbol n) = \{000\}$ only a single $I_{\pi \pi \pi}=3$ and a single $I_{\pi \pi \pi}=1$ state arise. The results of this exercise are given in the rightmost four columns of tables~\ref{tab:counting_ni_pipipi_P000},~\ref{tab:counting_ni_pipipi_P001}, and~\ref{tab:counting_ni_pipipi_P011}.

Finally, it is useful to project the states to definite finite-volume irreps. To explain this in some detail we let $G$ represent the octahedral group $O_h$ or a little group defined by the set of elements that leave the total momentum $\boldsymbol P$ invariant.
Selecting a given set of states with definite $S(\boldsymbol n)$ and a definite value of $I_{\pi \pi \pi}$, we define a vector space $V$. This leads to a (generally reducible) representation of $G$ on $V$, called $\rho$. (One can write $\rho: G \to \mathrm{GL}(V)$, where $\mathrm{GL}(V)$ is the general linear group of the vector space $V$.) The representation is unitary, also for the extension to higher spin particles, and for the present spin-zero case, the representation is additionally real and orthogonal.

As a specific example take $S(\boldsymbol n) = \{011\}$ and $I_{\pi \pi \pi} = 2$. The corresponding vector space $V$ is spanned by 6 vectors. These are constructed by assigning the $18$ distinct rotations and permutations of $\boldsymbol n_1 = (0,0,0), \boldsymbol n_2 = (0,0,1), \boldsymbol n_3 = (0,0,-1)$ to $\boldsymbol V$ defined in eq.~\eqref{eq:V_vector}, leading to 21 distinct states in the basis with definite individual pion flavors. Then projecting to $I_{\pi \pi \pi} = 2$ isolates 6 of the states defining $V$. Every element of $O_h$ acts physically on the momentum assignments such that each of the elements of $V$ is mapped to a linear combination of the others. This defines the representation $\rho$ of $O_h$ on $V$, concretely as a set of 48 6-dimensional matrices.

Next, we fix an irreducible unitary representation $\Gamma$ of $G$ with dimension $d_\Gamma$. For example, in the case of $\boldsymbol P = [000] \Leftrightarrow G = O_h$, $\Gamma$ is drawn from $A_1^+, A_1^-, A_2^+, A_2^-, E^+, E^-, T_1^+, T_1^-, T_2^+, T_2^-$ and $d_\Gamma$ is either $1$, $2$, or $3$. Now let $D^{(\Gamma)}(g)$ denote a $d_\Gamma \times d_\Gamma$ matrix for the irrep of group-element $g$ in some fixed orthonormal basis, and let $\chi_\Gamma(g) = \mathrm{tr}\, D^{(\Gamma)}(g)$ denote its character.
Then, we can define the standard character projector
\begin{equation}
P_\Gamma = \frac{d_\Gamma}{|G|}\,\sum_{g\in G} \chi_\Gamma(g^{-1})\,\rho(g) \,,
\end{equation}
where we have used $|G|$ as the order of $G$. This is an orthogonal projector satisfying $P_\Gamma^2=P_\Gamma$. For unitary $\Gamma$ we can equivalently write,
\begin{equation}
P_\Gamma = \frac{d_\Gamma}{|G|}\,\sum_{g\in G} \chi_\Gamma(g)^* \,\rho(g)\,,
\end{equation}
where the $*$ denotes complex conjugation.

We emphasize that this projector will project onto all rows of $\Gamma$. So if we let \(m_\Gamma\) be the multiplicity of \(\Gamma\) in \(\rho\) then $P_\Gamma$ will have $m_\Gamma \times d_\Gamma$ non-zero eigenvalues, each equal to 1.
(More technically we say that the image of $P_\Gamma$ is equal to the $\Gamma$-isotypic component of $V$, which is isomorphic to $\mathbb{C}^{m_\Gamma}\otimes \mathbb{C}^{d_\Gamma}$.)

This redundancy in the projection is not of essential importance when counting the non-interacting energies. It is more relevant, however, when projecting the matrices of the quantization condition itself, as these should be taken as small as possible to reduce computational cost. This is also relevant for the process of root finding. If the character projector is used to project the matrices in \Eq~\eqref{eq:QC_3}, then near a solution $E_n^{[I_{\pi \pi \pi}]}(L, \boldsymbol P, \boldsymbol \eta)$, the determinant will scale as $(E - E_n^{[I_{\pi \pi \pi}]}(L, \boldsymbol P, \boldsymbol \eta))^{d_\Gamma}$ leading to quantitatively different behavior, especially for even vs.~odd $d_\Gamma$.

Considering these points, it is convenient to instead define a row-specific projector.
We can readily do so by using matrix elements of $D^{(\Gamma)}(g)$ to define
\begin{equation}
P^{(\Gamma)}_{\mu\nu} = \frac{d_\Gamma}{|G|}\,\sum_{g\in G} \big(D^{(\Gamma)}_{\mu\nu}(g)\big)^*\,\rho(g) \,,
\end{equation}
which satisfy a generalized orthogonality relation, $P^{(\Gamma)}_{\mu\nu}\,P^{(\Gamma)}_{\alpha\beta}=\delta_{\nu\alpha}\,P^{(\Gamma)}_{\mu\beta}$.
In particular, if we restrict attention to the first row then we only need
\begin{equation}
P^{(\Gamma)}_{11} = \frac{d_\Gamma}{|G|}\,\sum_{g\in G} \big(D^{(\Gamma)}_{11}(g)\big)^*\,\rho(g) \,.
\end{equation}
This serves as an orthogonal projector onto the desired subspace and $P^{(\Gamma)}_{11}$ will only have $m_\Gamma$ non-zero eigenvalues, each equal to 1. (More technically the image of $P^{(\Gamma)}_{11}$ is isomorphic to $\mathbb{C}^{m_\Gamma}\otimes \mathrm{span}\{e_1\}$, where $e_1$ is the first basis vector in the representation space of $\Gamma$.) Moreover, $P_\Gamma = \sum_{\mu=1}^{d_\Gamma} P^{(\Gamma)}_{\mu\mu}$, so it is manifest that the character projector can be recoverd by summing up the row projectors.

Returning to our example with $S(\boldsymbol{n}) = \{011\}$ and $I_{\pi \pi \pi} = 2$, we can now apply the above procedure. We find that, for all irreps besides $A_1^-, E^-, T_1^+$, both the character and the row-specific projectors have only vanishing eigenvalues. Thus, only these three irreps appear in the decomposition, each with $m_\Gamma = 1$. Summing the dimensions of the contributing irreps gives $1+2+3=6$, which matches the counting established above.
This procedure is automated by the \ampyl\ package~\cite{ampyL}. The results for the lowest lying non-interacting states, for $\boldsymbol P = [000]$, $[001]$, $ [011]$ are summarized in tables~\ref{tab:ni_pipipi_P000_irreps}, \ref{tab:ni_pipipi_P001_irreps}, and \ref{tab:ni_pipipi_P011_irreps}, respectively.

Finally, as we motivated at the beginning of this discussion, the same procedure can be applied to the quantization condition itself, in particular to the matrices appearing in eq.~\eqref{eq:QC_3}, namely $\textbf K^{[I_{\pi \pi \pi}]}$, $\textbf F^{[I_{\pi \pi \pi}]}$, and $\textbf G^{[I_{\pi \pi \pi}]}$. The only distinction is that the vector space $V$ is now simply the space that we have already defined, with indices $f, \boldsymbol k, \ell, m$. The representation $\rho$ is then defined by the action of the relevant finite-volume symmetry group on the spectator's plane-wave momentum $\boldsymbol k$ as well as the dimer's angular momentum $\ell, m$. All other steps follow as above to construct the row-specific projector $P^{(\Gamma)}_{\mu\nu}$. Again, the $\ampyl$ package automates this process, delivering the quantization condition for a specific finite-volume irrep~\cite{ampyL}.

\begingroup
\renewcommand{\arraystretch}{1.3}
\begin{landscape}
\begin{table}[]
\centering
\begin{tabular}{c|c|c|c|c|c}
\hline\hline
\multirow{2}{*}{$S(\boldsymbol{n})$ }& \multicolumn{4}{c|}{$I_{\pi\pi\pi}$}& \multirow{2}{*}{dim.}\\
\cline{2-5}
& 3 & 2 & 1 & 0 & \\
\hline \hline
$\{000\}$ &$A_1^-$ & --- &$A_1^-$ & --- & 2\\
$\{011\}$ & $A_1^-, E^-$ & $A_1^-, E^-, T_1^+$ &$2A_1^-, 2E^-, T_1^+$ & $T_1^+$& 21 \\
$\{022\}$ & $A_1^-, E^-, T_2^-$ & $A_1^-, E^-, T_2^-, T_1^+, T_2^+$ & $2A_1^-, 2E^-, 2T_2^-, T_1^+, T_2^+$ & $T_1^+, T_2^+$ & 42 \\
$\{112\}$ & $A_1^-, E^-, T_2^-, T_1^+, T_2^+$ & $A_1^-, A_2^-, 2E^-, T_1^-, T_2^-, 2T_1^+, 2T_2^+$ & $2A_1^-, A_2^-, 3E^-, T_1^-, 2T_2^-, 3T_1^+, 3T_2^+$ & $A_2^-, E^-, T_1^-, T_1^+, T_2^+$ & 84 \\
$\{033\}$ & $A_1^-, T_2^-$ & $A_1^-, T_2^-, A_2^+, T_1^+$ & $2A_1^-, 2T_2^-, A_2^+, T_1^+$ & $A_2^+, T_1^+$ & 28\\
\hline\hline
\end{tabular}
\caption{Distribution of the lowest-lying $\boldsymbol{P} = [000]$ non-interacting states among $O_h$ irreps for each three-pion isospin channel. The intrinsic negative parity of the pion is included in these labels. The multiplier preceding a given irrep denotes its degeneracy within the specified set, and the final column lists the total number of states. These results reproduce table~5 of ref.~\cite{Hansen:2020zhy}.}
\label{tab:ni_pipipi_P000_irreps}
\end{table}

\begin{table}[]
\centering
\begin{tabular}{c|c|c|c|c|c}
\hline\hline
\multirow{2}{*}{$S(\boldsymbol{n})$ }& \multicolumn{4}{c|}{$I_{\pi\pi\pi}$}& \multirow{2}{*}{dim.}\\
\cline{2-5}
& 3 & 2 & 1 & 0 & \\ \hline\hline
$\{001\}$ & $A_2$ & $A_2$ & $2A_2$ & --- & 4\\
$\{012\}$ &$A_2, B_2, E_2$ & $2A_2, 2B_2, 2E_2$ & $3A_2, 3B_2, 3E_2$ & $A_2, B_2, E_2$ & 28\\
$\{111\}$ & $A_2, B_2$ & $A_2, B_2, E_2$ & $ 2A_2, 2B_2, E_2$ & $E_2$ & 14\\
$\{111\}^*$ & $A_2$ & $A_2$ & $2A_2$ & --- & 4\\
$\{014\}$ & $A_2$ & $2A_2$ & $3A_2$ & $A_2$ &7 \\
$\{023\}$ & $A_2, B_1, E_2$ & $2A_2, 2B_1, 2E_2$ & $3A_2, 3B_1, 3E_2$ & $A_2, B_1, E_2$ & 28\\
$\{113\}$ & $A_2, B_1, E_2$ & $A_2, A_1, B_2, B_1, 2E_2$ & $2A_2, A_1, B_2, 2B_1, 3E_2$ & $A_1, B_2, E_2$ & 28 \\
\hline\hline
\end{tabular}
\caption{Distribution of the lowest-lying $\boldsymbol{P} = [001]$ non-interacting states among ${\rm Dic}_4$ irreps for each three-pion isospin channel, with other details as in table~\ref{tab:ni_pipipi_P000_irreps}. Here $\{111\}$ refers to $\boldsymbol n_1 = (0,1,0), \boldsymbol n_2 = (0,-1,0), \boldsymbol n_3 = (0,0,1)$ while $\{111\}^*$ refers to $\boldsymbol n_1 = (0,0,1), \boldsymbol n_2 = (0,0,1), \boldsymbol n_3 = (0,0,-1)$.}
\label{tab:ni_pipipi_P001_irreps}
\end{table}
\end{landscape}
\renewcommand{\arraystretch}{1}
\endgroup

\begingroup
\renewcommand{\arraystretch}{1.2}
\begin{table}[]
\centering
\begin{tabular}{c|c|c|c|c|c}
\hline
\multirow{2}{*}{$S(\boldsymbol{n})$ }& \multicolumn{4}{c|}{$I_{\pi\pi\pi}$}& \multirow{2}{*}{dim.}\\
\cline{2-5}
& 3 & 2 & 1 & 0 & \\ \hline\hline
$\{002\}$ & $A_2$ & $A_2$ & $2A_2$ & --- & 4\\
$\{011\}$ & $A_2$ & $A_2, B_2$ & $2A_2, B_2$ & $B_2$ & 7 \\
$\{012\}$ & $A_2, B_1$ & $2A_2, 2B_1$ & $3A_2, 3B_1$ & $A_2, B_1$ & 14 \\
$\{022\}$ & $A_2, A_1$ & $A_2, A_1, B_2, B_1$ & $2A_2, 2A_1, B_2, B_1$ & $B_2, B_1$ & 14\\
$\{112\}$ & $A_2, A_1, B_2, B_1$ & $2A_2, 2A_1, 2B_2, 2B_1$ & $3A_2, 3A_1, 3B_2, 3B_1$ & $A_2, A_1, B_2, B_1$ & 28 \\
\hline\hline
\end{tabular}
\caption{Distribution of the lowest-lying $\boldsymbol{P} = [011]$ non-interacting states among ${\rm Dic}_2$ irreps for each three-pion isospin channel ($I_{\pi\pi\pi} = 3, 2, 1,$ and $0$), with other details as in table~\ref{tab:ni_pipipi_P000_irreps}.}
\label{tab:ni_pipipi_P011_irreps}
\end{table}
\renewcommand{\arraystretch}{1}
\endgroup

\subsection{K-matrix parametrizations}
\label{sec:Kmatrix_param}

As discussed in section~\ref{sec:Kmatrix}, the quantization condition allows one to predict finite-volume energies for a given truncation and parametrization of the two- and three-particle K-matrices. In this work we set $\mathcal K_{{\rm df},3} = 0$ so that it only remains to parametrize the two-body dynamics, and to make a specific choice for the smooth cutoff $H(E_{2,\boldsymbol{k}}^{\star 2})$.

We restrict attention to the simplest truncation: $\ell_{\sf max}^{\{2\}} = 0$, $\ell_{\sf max}^{\{1\}} = 1$, and $\ell_{\sf max}^{\{0\}} = 0$, implying that only one non-zero phase shift is included in each channel. We then set the parametrizations as
\begin{align}
\delta_0^{\{2\}}(p) & = \delta^{{\sf SL} ,\{2\}}_{0}(p \, \vert \, \boldsymbol \eta^{{\sf SL},\{2\}}_{0}) \,,
\\
\delta_1^{\{1\}}(p) & = \delta^{{\sf BW} ,\{1\}}_{1}(p \, \vert \, \boldsymbol \eta^{{\sf BW},\{1\}}_{1}) \,,
\\
\delta_0^{\{0\}}(p) & = \delta^{{\sf BW} ,\{0\}}_{0}(p \, \vert \, \boldsymbol \eta^{{\sf BW},\{0\}}_{0}) \,,
\end{align}
where ${\sf SL}$ stands for the \emph{scattering length} and ${\sf BW}$ stands for \emph{Breit--Wigner}.

In words, this means that the isospin-two case, in which no resonance appears, is described with a single scattering length while the isospin-one and isospin-zero cases are described with Breit--Wigners, for the $\rho$ and $\sigma$ resonances respectively.%
\footnote{We note that Breit--Wigner representation does not provide a good description for the $\sigma$ resonance at physical masses. However, this is not a concern in this context as it is used to demonstrate the implementation of the formalism. Additionally, the Breit--Wigner representation is descriptive for a range of heavier-than-physical pion masses for which the $\sigma$ pole is closer to the real axis~\cite{Rodas:2023gma, Rodas:2023nec,Briceno:2017qmb, Briceno:2016mjc}.} It follows that a total of five parameters are needed to describe the three K-matrices:
\begin{equation}
\boldsymbol \eta^{{\sf SL},\{2\}}_{0} = \{a_{0}\} \,, \qquad \boldsymbol \eta^{{\sf BW},\{1\}}_{1} = \{m_{\rho}, g_{\rho}\} \,, \qquad \boldsymbol \eta^{{\sf BW},\{0\}}_{0} = \{m_{\sigma}, g_{\sigma}\} \,.
\end{equation}
The explicit functional forms are
\begin{align}
p \cot \delta^{{\sf SL} ,\{2\}}_{0}(p \, \vert \, a_{0}) & = - \frac{1}{a_0} \,,
\\ \label{eq:iso1_param}
p^3 \cot \delta^{{\sf BW} ,\{1\}}_{1}(p \, \vert \, m_{\rho}, g_{\rho}) & = \frac{6\pi}{g_\rho^2} E (m_\rho^2 - E^2) \,,
\\ \label{eq:iso0_param}
p \cot \delta^{{\sf BW} ,\{0\}}_{0}(p \, \vert \, m_{\sigma}, g_{\sigma}) & = \frac{6\pi}{g_\sigma^2} \frac{m_\sigma^2 - E^2}{E }
\,.
\end{align}

It remains only need to consider $H(E_{2,\boldsymbol{k}}^{\star 2})$. In section~\ref{sec:Kmatrix}, we have already noted that a one-parameter family of cutoff functions can be used to adjust the point below which the smooth cutoff function has zero support, as a function of the dimer CMF energy. This is useful for removing dependence on a region where a particular parameterization of the two-particle K matrices leads to unphysical behavior. In the present case, the issue is that deep subthreshold poles appear in the Breit--Wigner parameterization for $E_{2,\boldsymbol k}^{\star 2} \ll 4 m_\pi^2$. We avoid this by taking $\alpha = -0.7$, which implies that the cutoff functions have zero support for $E_{2,\boldsymbol k}^{\star 2} < 0.3\, m_\pi^2$, and thus the unphysical pole is never encountered. For simplicity, we make this choice for all cutoff functions, although, strictly speaking, it is only necessary for those two-pion isospin channels in which the unphysical poles arise.

Given these details, the quantization condition is fully defined and can be evaluated for any given set of parameters.

\section{Results}
\label{sec:results}

We turn now to our numerical results for the finite-volume energies for various values of the box size $L$, the total isospin $I_{\pi \pi \pi}$, the total momentum in the finite-volume frame $\boldsymbol P$, the finite-volume irrep, and the K-matrix parameters $\boldsymbol \eta$. This section is divided into subsections, corresponding to the three values of isospin that we consider. In all extractions, the finite-volume energies are presented in a range of $L$ corresponding to $3.8 < m_\pi L < 6.0$.

Root-finding the quantization condition is delicate due to the high density of solutions in certain energy regions and the fact that $\mathcal Q^{[I_{\pi \pi \pi}]}(E, \boldsymbol P, L \, \vert \, \boldsymbol \eta) $ contains numerous poles in addition to zero crossings. This issue is largely circumvented by starting with the non-interacting energies, projected to the relevant irreps, and then setting weakly interacting K-matrices while searching for solutions near the non-interacting energies.

In the case where two-particle sub-process resonances appear, it is useful to set the widths (equivalently the couplings) of the resonances to be artificially small while keeping the mass parameters at their target values. Then the energies associated with a resonance and spectator can be found, with expectations set by the non-interacting energies. Once the spectrum is known for the artificially weak interactions, one can adiabatically increase to the target parameters, finding roots at each step with a search range motivated by the previous solution. This ensures that no energies are missed and that the correct spectrum is determined in each case. In certain cases, we supplement this strategy with a dense scan of $\mathcal Q^{[I_{\pi \pi \pi}]}(E, \boldsymbol P, L \, \vert \, \boldsymbol \eta)$, to determine whether any additional solutions are present.

In addition to the three-pion states shown in tables~\ref{tab:ni_pipipi_P000_irreps}, \ref{tab:ni_pipipi_P001_irreps}, and \ref{tab:ni_pipipi_P011_irreps}, we list the lowest non-interacting energy states of $\rho\pi$ and $\sigma\pi$ (for the case of a stable $\rho$ and a stable $\sigma$) in tables~\ref{tab:ni_rhopi_P000_P001_irreps} and \ref{tab:ni_sigpi_P000_irreps} respectively.

\begingroup
\renewcommand{\arraystretch}{1.3}
\begin{table}[h]
\centering
\begin{tabular}{c|c|c|c}
&$\{\boldsymbol n^2_\rho \boldsymbol n^2_\pi \}$& irrep & dim.\\
\hline \hline
\multirow[c]{5}{*}{\rotatebox{90}{$\boldsymbol{P} = [000]$}} & $\{00\}$ & $T_1^+$ & 3 \\
&$\{11\}$ & $A_1^-, E^-, T_1^-, T_2^-, 2T_1^+, T_2^+$ & 18\\
&$\{22\}$ & $A_1^-, A_2^-, 2E^-, 2T_1^-, 2T_2^-, A_2^+, E^+, 3T_1^+, 2T_2^+$& 36 \\
& $\{33\}$ & $A_1^-, E^-, T_1^-, 2T_2^-, A_2^+, E^+, 2T_1^+, T_2^+$ & 24\\
&$\{44\}$ & $A_1^-, E^-, T_1^-, T_2^-, 2T_1^+, T_2^+$ & 18 \\
\hline \hline
\multirow[c]{6}{*}{\rotatebox{90}{$\boldsymbol{P} = [001]$}} & $\{10\}$ & $A_2, E_2$ & 3 \\
& $\{01\}$ & $A_2, E_2$ & 3\\
& $\{21\}$ & $2A_2, A_1, 2B_2, B_1, 3E_2$ & 12 \\
& $\{12\}$ & $2A_2, A_1, 2B_2, B_1, 3E_2$ & 12 \\
& $\{41\}$ & $A_2, E_2$ & 3\\
& $\{32\}$ & $2A_2, A_1, B_2, 2B_1, 3E_2$ & 12 \\
\hline\hline
\end{tabular}
\caption{Distribution of the lowest-lying $\boldsymbol{P} = [000]$ and $\boldsymbol{P} = [001]$ non-interacting $\rho \pi$ states among $O_h$ and ${\rm Dic}_4$ irreps, respectively. States are labeled by the squared dimensionless momenta of the $\rho$ and $\pi$. The listed irreps appear for each allowed isospin ($I_{\rho\pi} = 2, 1,$ and $0$). Other details are as in table~\ref{tab:ni_pipipi_P000_irreps}.}
\label{tab:ni_rhopi_P000_P001_irreps}
\end{table}

\begin{table}[h]
\centering
\begin{tabular}{c|c|c}
$\{\boldsymbol n_\sigma^2 \boldsymbol n_\pi^2 \}$ & irrep & dim. \\
\hline\hline
$\{00\}$ & $A_1^-$ & 1 \\
$\{11\}$ & $A_1^-, E^-, T_1^+$& 6\\
$\{22\}$ & $A_1^-, E^-, T_2^-, T_1^+, T_2^+$ & 12 \\
$\{33\}$ & $A_1^-, T_2^-, A_2^+, T_1^+$ & 8 \\
\hline\hline
\end{tabular}
\caption{Distribution of the lowest-lying $\boldsymbol{P} = [000]$ non-interacting $\sigma \pi$ states among $O_h$ irreps, with other details as for table~\ref{tab:ni_pipipi_P000_irreps}.}
\label{tab:ni_sigpi_P000_irreps}
\end{table}
\renewcommand{\arraystretch}{1}
\endgroup

\subsection{\texorpdfstring{$I_{\pi\pi\pi}=2$}{I(pipipi)=2}}
\label{sec:I2}

\begin{figure}
\centering
\includegraphics[width=\textwidth]{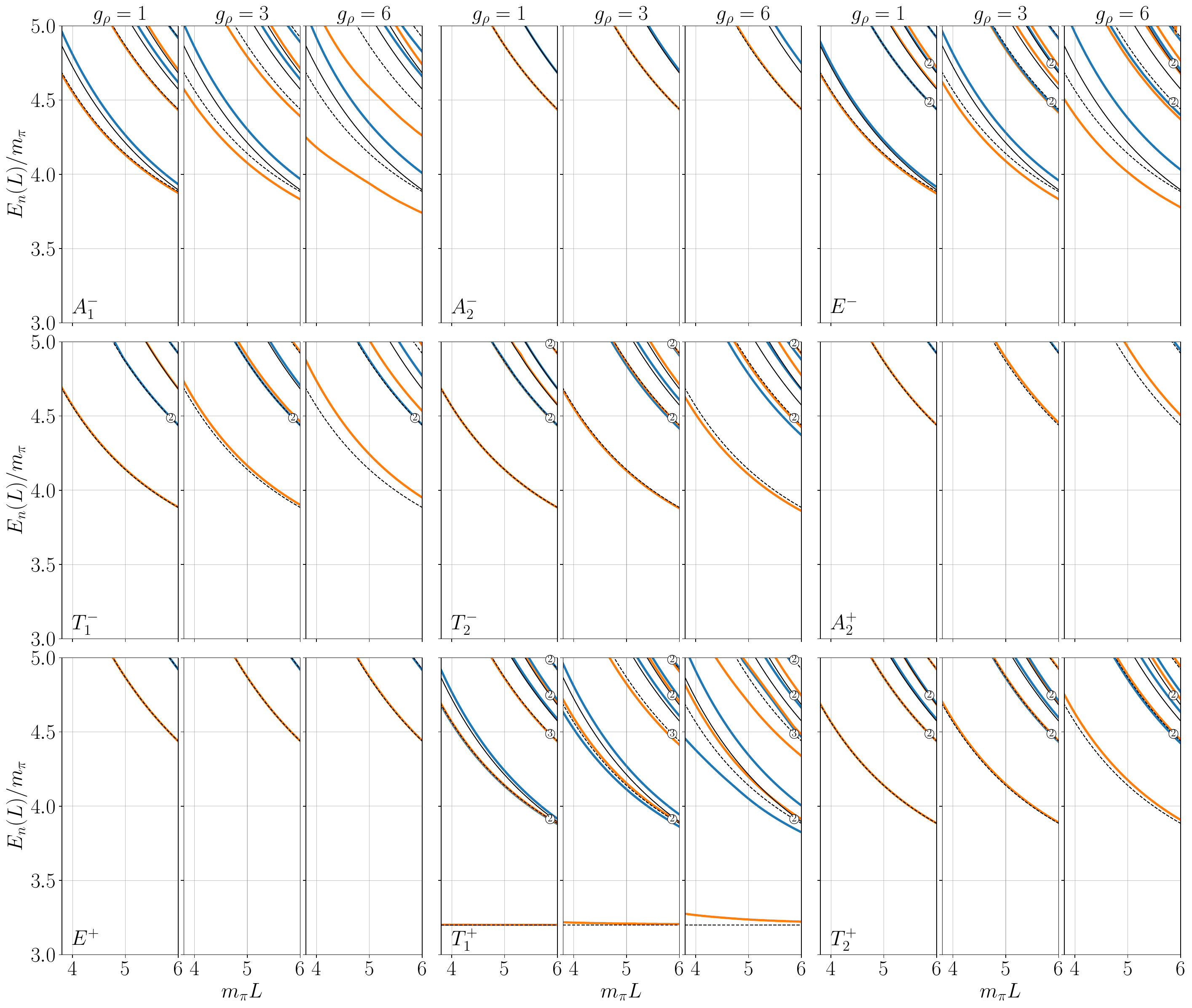}
\caption{Finite-volume energy spectra $E_n(L)$ as functions of the box size $L$ for isospin $I_{\pi\pi\pi} = 2$ and total momentum $\boldsymbol{P} = [000]$. Alternating colors distinguish neighboring curves. As described in the text, the three-particle interaction is set to zero ($\mathcal{K}_{{\rm df},3} = 0$), leaving two distinct two-particle K-matrices to parametrize. We take a scattering-length-only description for the $I_{\pi\pi} = 2$ subsystem (with $m a_0 = 0.1$), and a Breit--Wigner description for the $I_{\pi\pi} = 1$ subsystem (with $m_\rho/m_\pi = 2.2$). The Breit--Wigner coupling varies across the three panels in each subfigure as indicated, and each of the nine subfigures corresponds to a finite-volume irrep, as labeled. Non-interacting energies are shown in black: dashed lines for $\rho\pi$ states and solid lines for $\pi\pi\pi$ states, with multiplicities indicated when greater than one. The non-interacting $\rho\pi$ energies are computed by treating the $\rho$ as a stable particle with the Breit--Wigner mass.}
\label{fig:iso2_nonmoving}
\end{figure}

We begin with \( I_{\pi\pi\pi} = 2 \), first focusing on energies for $\boldsymbol P = [000]$, shown in \Fig~\ref{fig:iso2_nonmoving}. As the $I_{\pi \pi} = 0$ subsystem does not contribute for $I_{\pi \pi \pi} = 2$, these spectra depend on three K-matrix parameters: $ a_0$, $m_\rho$, and $g_\rho$. In each figure, we present the results for $m_\pi a_0 = 0.1$ and $m_\rho/m_\pi = 2.2$ while $g_\rho$ is varied as shown in the figure labels. We present spectra for nine of the ten irreps of \(O_h\), with interacting energies (orange and blue) compared to non-interacting levels (black). The latter are labeled by their multiplicity and solid (dashed) curves indicate \(\pi\pi\pi\) (\(\rho\pi\)) levels. We stress that $\rho \pi$ levels here refer to the exercise of keeping $m_\rho$ fixed while reducing $g_\rho$ to zero.

The \( A_1^+ \) spectrum is omitted here as its lowest lying energy appears for \(\rho\pi\) states at \{5,5\} and for \(\pi\pi\pi\) at \{6,2,2\}. By contrast, the most populated irrep, \( T_1^+ \), contains the lowest three \( \rho\pi \) energy states and the next-to-lowest four \( \pi\pi\pi \) states, consistent with the levels shown in tables~\ref{tab:ni_pipipi_P000_irreps} and \ref{tab:ni_rhopi_P000_P001_irreps}.

As is clear from the figures, interacting energies deviate from non-interacting ones by shifts that are power-like in $1/L$. The interactions also lift degeneracies, splitting energy levels according to the multiplicities shown. In each case, one can confirm that the total count of interacting and non-interacting levels matches when multiplicities are considered.

\begin{figure}[h]
\centering
\includegraphics[width=\textwidth]{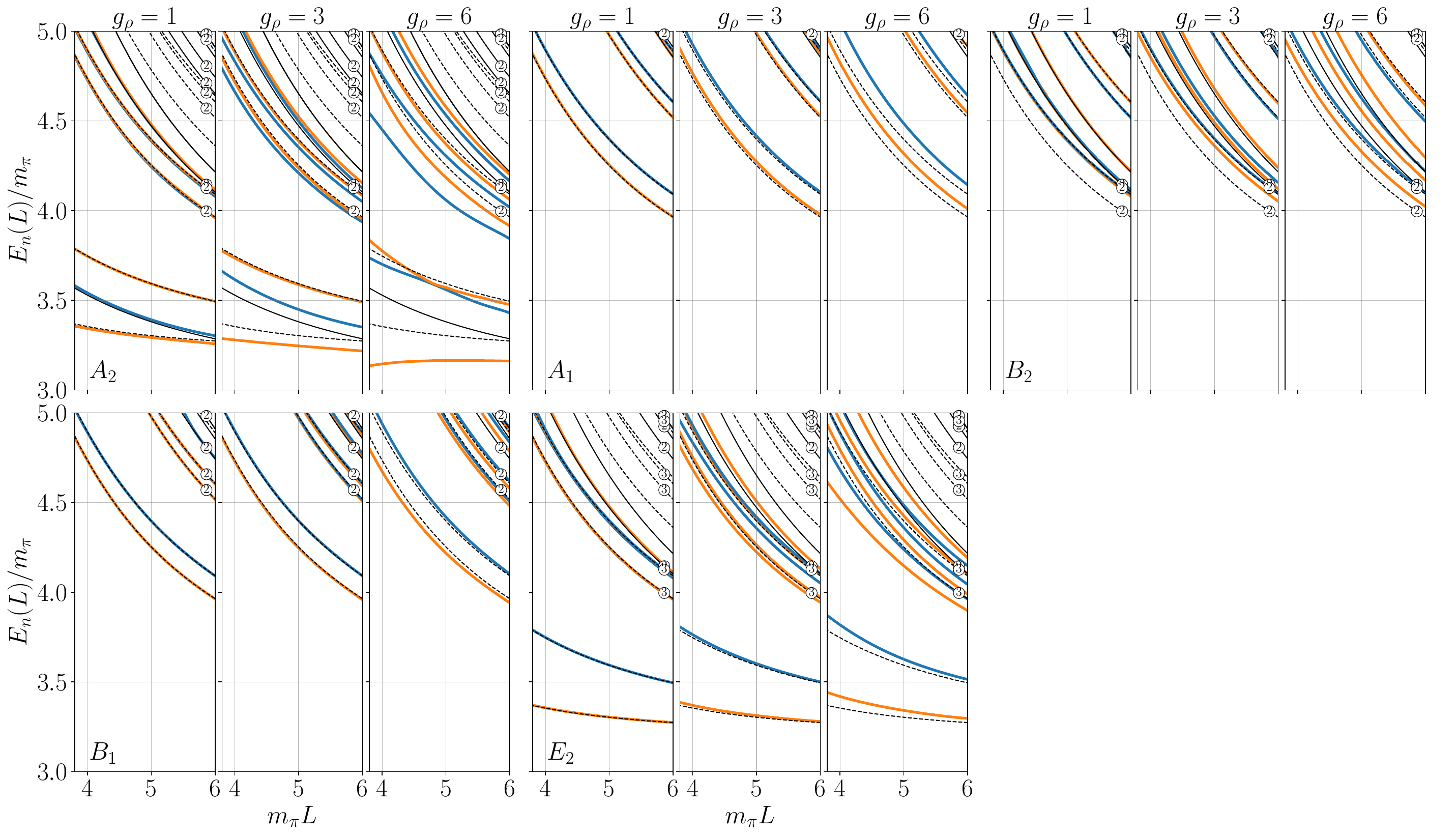}
\caption{Finite-volume energy spectra $E_n(L)$ as functions of the box size $L$ for isospin $I_{\pi\pi\pi} = 2$ and total momentum $\boldsymbol{P} = [001]$. Other details as in figure~\ref{fig:iso2_nonmoving}. In some cases only the lowest lying $n$ interacting states are shown (for some positive integer $n$) as the high density of states makes the remaining excited states uninformative.}
\label{fig:iso2_moving}
\end{figure}

In \Fig~\ref{fig:iso2_moving} we turn to the quantization condition in a moving frame with a total momentum $\boldsymbol{P} = [001]$. In this case, the symmetry group is Dic$_4$ and one identifies five irreps, as listed in table~\ref{tab:irrep_review}. As such, the density of states is generically expected to be higher in a given irrep as seen in the figure. The $A_2$ irrep contains the lowest energy states of $\rho\pi$ and $\pi\pi\pi$ listed in tables~\ref{tab:ni_pipipi_P001_irreps} and \ref{tab:ni_rhopi_P000_P001_irreps} for $\boldsymbol{P} = [001]$. Again, we have confirmed that the level counting connects with the non-interacting states as expected.

\subsection{\texorpdfstring{$I_{\pi\pi\pi}=1$}{I(pipipi)=1}}
\label{sec:I1}

\begin{figure}
\centering
\includegraphics[width=\textwidth]{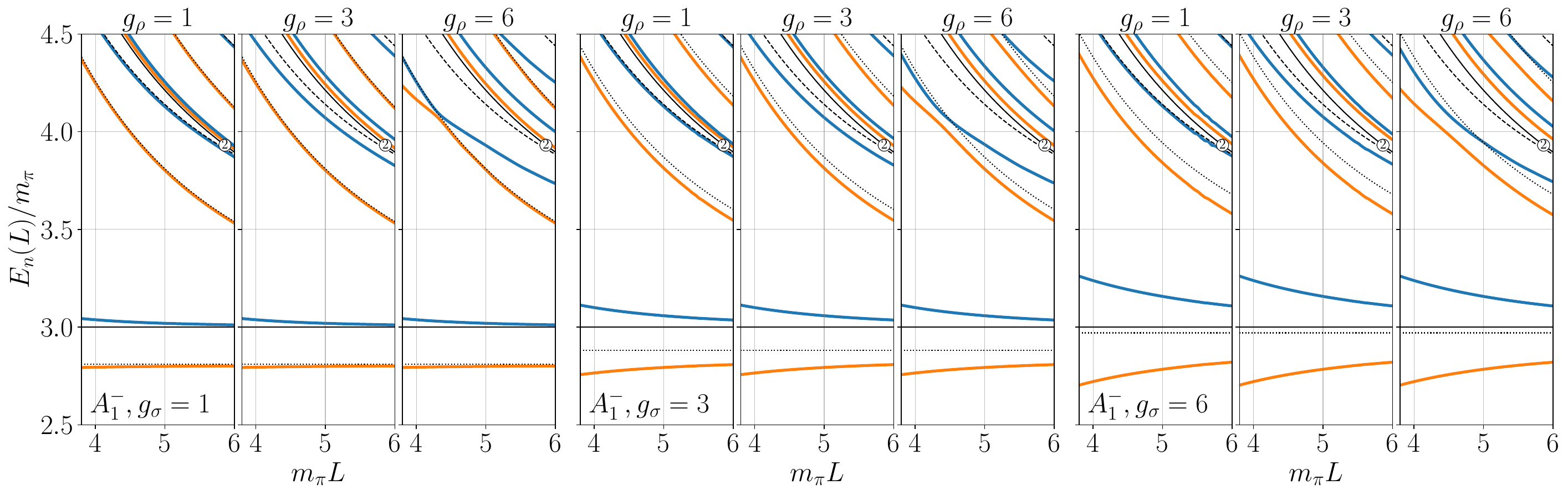}
\caption{
Finite-volume energy spectra, $E_n(L)$, as functions of the box size $L$ for isospin $I_{\pi\pi\pi} = 1$ and total momentum $\boldsymbol{P} = [000]$. Alternating colors distinguish neighboring curves. Results are shown for a single finite-volume irrep, as labeled. The $I_{\pi\pi} = 2$ subsystem is described using only a scattering length ($m_\pi a_0 = 0.1$), while the $I_{\pi\pi} = 1$ and $I_{\pi\pi} = 0$ subsystems use Breit--Wigner parameterizations with $m_\rho/m_\pi = 2.2$ and $m_\sigma/m_\pi = 1.8$, respectively. The $I_{\pi\pi} = 1$ Breit--Wigner coupling ($g_\rho$) varies across the three panels within each subfigure, while the $I_{\pi\pi} = 0$ coupling ($g_\sigma$) varies across the three subfigures, as indicated. Non-interacting energies are shown in black: dotted lines for $\sigma\pi$ states, dashed lines for $\rho\pi$ states, and solid lines for $\pi\pi\pi$ states, with multiplicities labeled when greater than one. Non-interacting $\sigma\pi$ energies use the $\sigma$ pole mass, while $\rho\pi$ energies use the $\rho$ Breit--Wigner mass.}
\label{fig:iso1_unmoving}
\end{figure}

Moving to the $I_{\pi\pi\pi} = 1$ channel, this is the most complex due to its coupling to all three two-pion isospins as shown in \Eq~\eqref{eq:iso1_vector}. We present the results for the same $I_{\pi \pi}=2$ and $I_{\pi \pi}=1$ parameters as above ($m_\pi a_0 = 0.1$ and $m_\rho/m_\pi = 2.2$). For the $I_{\pi \pi} = 0$ channel that appears here for the first time, we take $m_\sigma/m_\pi = 1.8$ so that the $\sigma$ forms a two-pion bound state. The two couplings $g_\rho$ and $g_\sigma$ are varied as shown in the figure labels.

Here we focus on extracting the finite-volume energies for a single one-dimensional irrep of the $\boldsymbol P = [000]$ system, namely the $A_1^-$. The spectra nicely display the intricacies of the coupled $\rho$ and $\sigma$ resonant channels. We highlight avoided level crossing between $\sigma\pi$ and $\rho\pi$ like levels.

\subsection{\texorpdfstring{$I_{\pi\pi\pi}=0$}{I(pipipi)=0}}

\begin{figure}
\centering
\includegraphics[width=\textwidth]{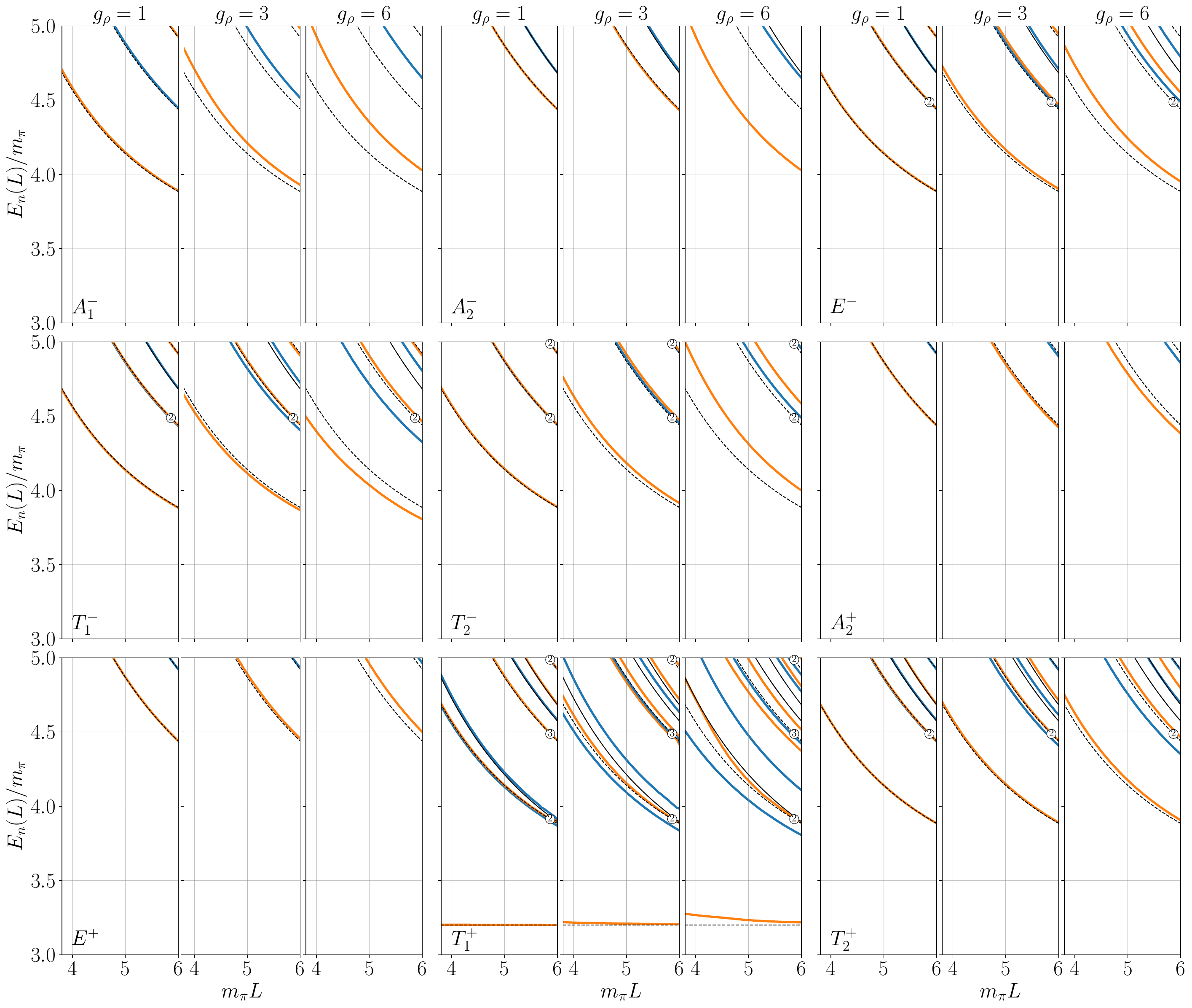}
\caption{Finite-volume energy spectra $E_n(L)$ as functions of the box size $L$ for isospin $I_{\pi\pi\pi} = 0$ and total momentum $\boldsymbol{P} = [000]$. Other details as in figure~\ref{fig:iso2_nonmoving}.}
\label{fig:iso0_unmoving}
\end{figure}

\begin{figure}[h]
\centering
\includegraphics[width=\textwidth]{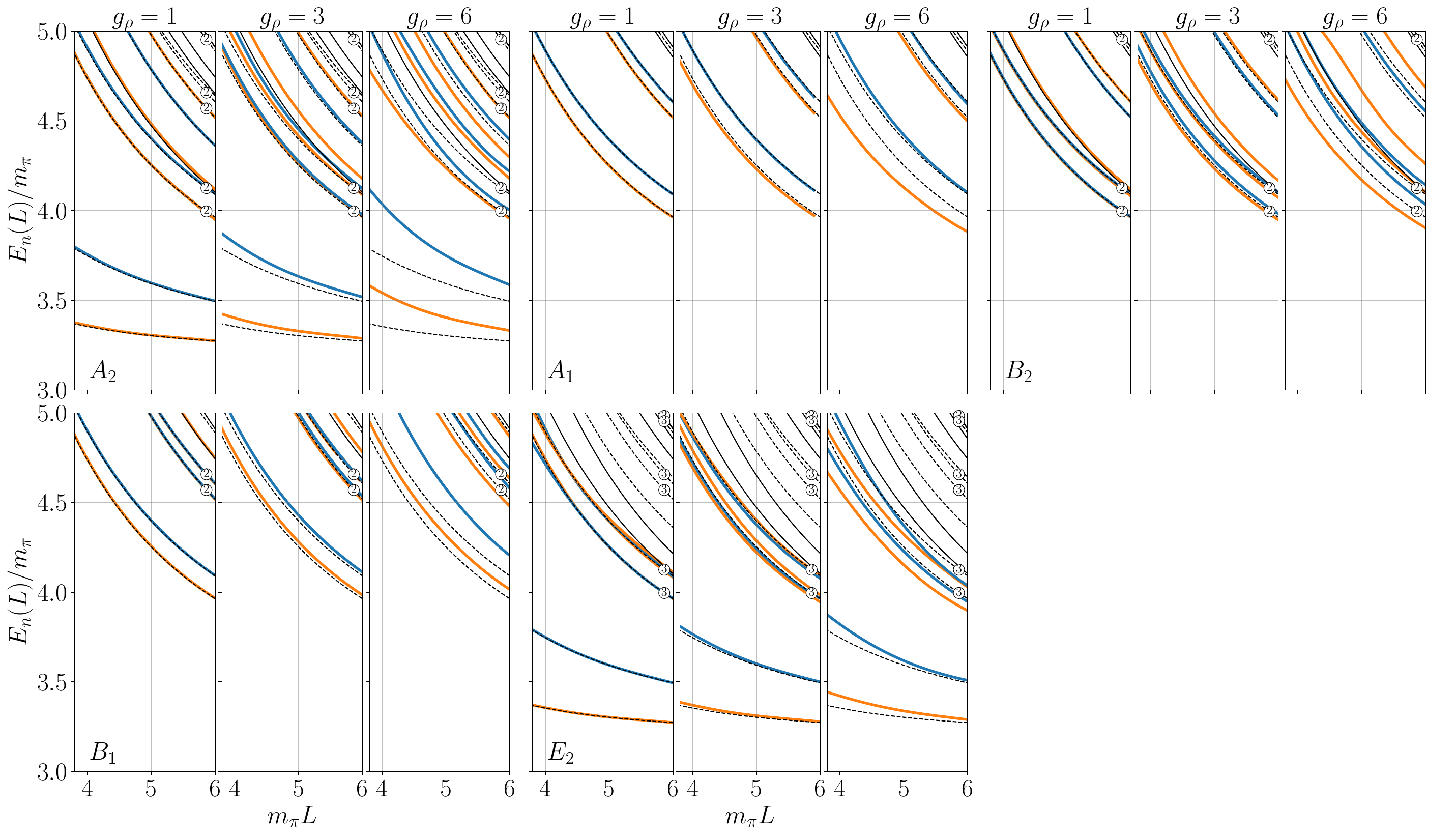}
\caption{Finite-volume energy spectra $E_n(L)$ as functions of the box size $L$ for isospin $I_{\pi\pi\pi} = 0$ and total momentum $\boldsymbol{P} = [001]$. Other details as in figure~\ref{fig:iso2_nonmoving}. In some cases only the lowest lying $n$ interacting states are shown (for some positive integer $n$) as the high density of states makes the remaining excited states uninformative.}
\label{fig:iso0_moving}
\end{figure}

Finally we turn to $I_{\pi\pi\pi}=0$. Although the simplest of the three isospin channels implemented in this work, this channel carries significant importance due to the presence of many three-particle resonances, such as $\omega(782)$ and $h_1(1170)$. Here only the $I_{\pi \pi}=1$ channel contributes and we continue to use the parameters established above ($m_\rho/m_\pi = 2.2$ with coupling $g_\rho$ varied). The finite-volume energies are shown in \Fig~\ref{fig:iso0_unmoving} for $\boldsymbol P = [000]$ and in \Fig~\ref{fig:iso0_moving} for $\boldsymbol P = [001]$.

\section{Conclusion}
\label{sec:conclusion}

In this paper we have demonstrated major progress in a fully implemented framework for studying three pions with general isospin using the RFT formalism of ref.~\cite{Hansen:2020zhy}. Here we have focused on the case of a vanishing short-distance three-body interactions ($\mathcal K_{\rm{df},3}=0$) so that these results provide a benchmark for future lattice calculations of three-pion systems, especially those where three-pion resonances are expected to appear. Multi-pion systems exhibit a rich phenomenology with resonances such as $\omega(782)$, $h_1(1170)$ and $a_1(1260)$. This work lays the groundwork for upcoming numerical lattice QCD calculations of the properties of such resonances. In fact full calculations are already progressing: an impressive recent example concerning the $\omega(782)$ resonance was presented in ref.~\cite{Yan:2024gwp}.

Extracting the energy spectrum from three-particle quantization conditions with coupled channels comes with a non-negligible computational cost, especially considering it is at the analysis stage of the calculation. This arises from the relatively large space of the matrices defining the quantization condition. In this work we have presented various steps towards accelerating this, many of which are contained in the open-source library \ampyl~\cite{ampyL}. This library is designed to facilitate the numerical evaluation of the quantization condition, to project matrices to row-specific finite-volume irreps, and to use knowlege of the non-interacting spectra to ensure all solutions are accounted for when root-finding the quantization condition.

One natural extension of this work would be efficient numerical evaluation of the Lellouch-L\"{u}scher factors for $K \rightarrow \pi\pi\pi$ weak decays, using the formalism of ref.~\cite{Hansen:2021ofl}. Among other things, this could provide basis of comparison for relating the RFT approach to other methods, e.g.~\Reference~\cite{pang2024lellouch}.

The primary focus moving forward will be the combination of the efficiently implemented formalism with numerical lattice QCD to produce fully systematic predictions of three-pion amplitudes and resonance properties. Building on the progress reported in refs.~\cite{Mai:2018djl,Horz:2019rrn,Blanton:2019vdk,Culver:2019vvu,Mai:2019fba,Fischer:2020jzp,Hansen:2020otl,Draper:2023boj,Dawid:2025zxc,Dawid:2025doq}, this will require reliably determining the finite-volume energy spectrum from lattice calculations and subsequently fitting the K-matrices using the relation between energies and the latter as presented in this work. The final stage---beyond the scope of the present study---involves applying integral equations to connect the resulting K-matrices to physical scattering amplitudes. For significant progress on the detailed workflow of this final step, see refs.~\cite{Jackura:2020bsk,Dawid:2023jrj,Jackura:2023qtp,Briceno:2024ehy,Jackura:2025wbw}.

\section*{Acknowledgements}

We thank Christopher Thomas for valuable discussions, particularly regarding finite-volume group theory. MTH also gratefully acknowledges assistance from Christopher Thomas with various cross-checks of the \ampyl\ library, especially those relevant to the code used to generate the tables in this work. We additionally thank Robert Edwards, Andrew Jackura, Fernando Romero-López, Steve Sharpe, Christopher Thomas, and David Wilson for past collaborations related to this work and for useful discussions. The work of Athari Alotaibi is funded by King Saud University (Riyadh, Saudi Arabia). RAB was partly supported by the U.S. Department of Energy, Office of Science, Office of Nuclear Physics under Award No. DE-SC0025665 and No. DE-AC02-05CH11231. MTH is supported in part by UK STFC grants ST/T000600/1 and ST/X000494/1 and additionally by UKRI Future Leader Fellowship MR/T019956/1.

\bibliographystyle{JHEP}
\bibliography{refs.bib}
\end{document}